\documentclass[sigconf]{acmart}
\usepackage{xcolor}
\usepackage{tabularx}
\usepackage{booktabs}
\usepackage{float}

\renewcommand\footnotetextcopyrightpermission[1]{}
\AtBeginDocument{%
  }

\begin{document}

\title{Toys that listen, talk, and play: Understanding Children's Sensemaking and Interactions with AI Toys}

\author{Aayushi Dangol$^*$}
\thanks{$^*$Both authors contributed equally.}
\affiliation{%
  \institution{University of Washington}
  \city{Seattle}
  \state{WA}
  \country{USA}
}

\author{Meghna Gupta$^*$}
\affiliation{%
  \institution{University of Washington}
  \city{Seattle}
  \country{USA}
}

\author{Daeun Yoo}
\affiliation{%
  \institution{University of Washington}
  \city{Seattle}
  \state{WA}
  \country{USA}
}

\author{Robert Wolfe}
\affiliation{%
  \institution{Rutgers University}
  \city{New Brunswick}
  \state{NJ}
  \country{USA}
}

\author{Jason Yip}
\affiliation{%
  \institution{University of Washington}
  \city{Seattle}
  \state{WA}
  \country{USA}
}

\author{Franziska Roesner}
\affiliation{%
  \institution{University of Washington}
  \city{Seattle}
  \state{WA}
  \country{USA}
}

\author{Julie A. Kientz}
\affiliation{%
  \institution{University of Washington}
  \city{Seattle}
  \state{WA}
  \country{USA}
}

\renewcommand{\shortauthors}{Dangol \& Gupta et al.}

\begin{abstract}
Generative AI (genAI) is increasingly being integrated into children’s everyday lives, not only through screens but also through so-called “screen-free” AI toys. These toys can simulate emotions, personalize responses, and recall prior interactions, creating the illusion of an ongoing social connection. Such capabilities raise important questions about how children understand boundaries, agency, and relationships when interacting with AI toys. To investigate this, we conducted two participatory design sessions with eight children ages 6 - 11 where they engaged with three different AI toys, shifting between play, experimentation, and reflection. Our findings reveal that children approached AI toys with genuine curiosity, profiling them as social beings. However, frequent interaction breakdowns and mismatches between apparent intelligence and toy-like form disrupted expectations around play and led to adversarial play. We conclude with implications and design provocations to navigate children’s encounters with AI toys in more transparent, developmentally appropriate, and responsible ways.
\end{abstract}



\begin{teaserfigure}
\centering
  \includegraphics[width=0.7\textwidth]{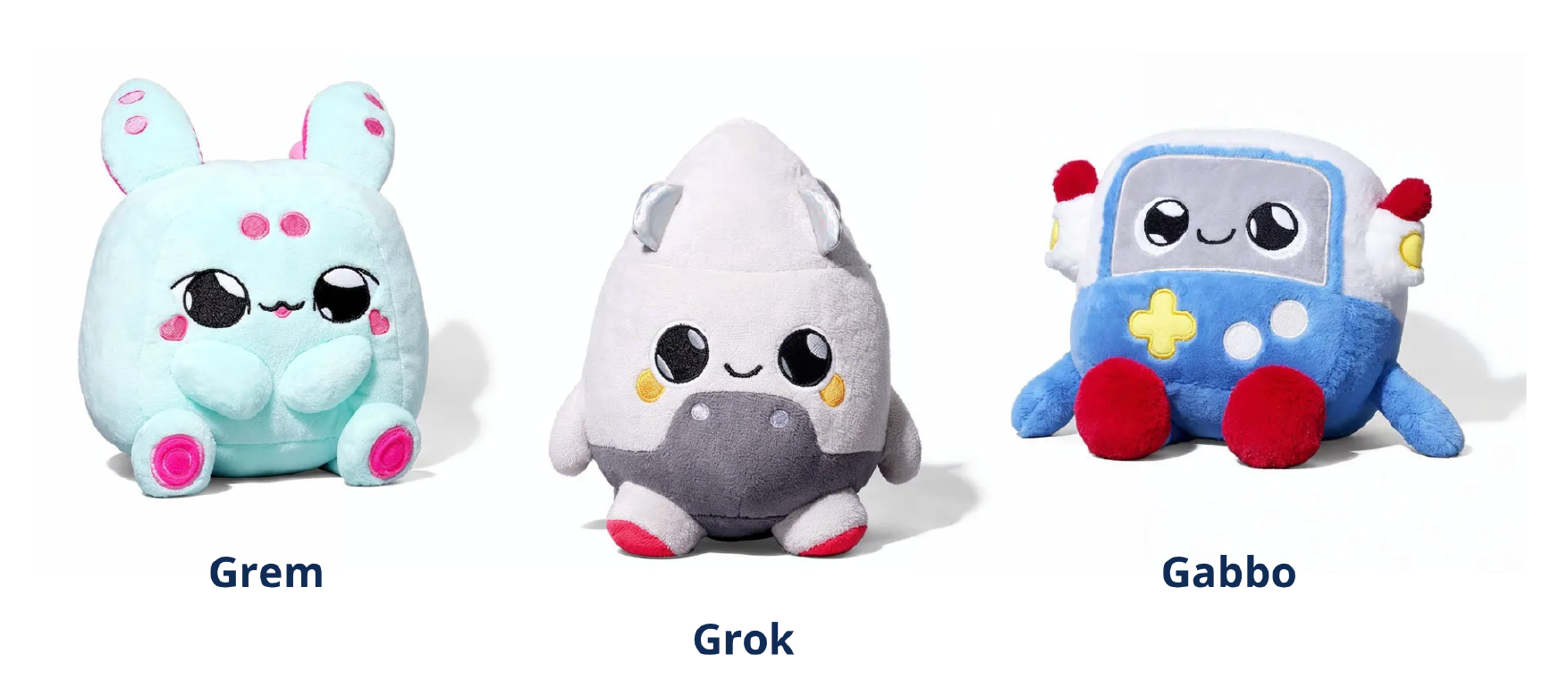}
  \caption{\textbf{Examples of emerging AI toys (Grem, Grok, and Gabbo) from https://heycurio.com/}}
  \label{fig:curio-toys}
\end{teaserfigure}

\maketitle

\section{Introduction}

Visions for how generative AI (genAI) could reshape children’s experiences are rapidly evolving, including the emergence of AI toys that can listen, talk, and respond dynamically during play~\cite{Wong2025AIToysTalkingBack, CommonSenseMedia2026AIToys, CrossErlich2025AIComesToPlaytime}. These systems are typically built on large language models (LLMs), originally developed for general-purpose use, and are marketed as offering playful, educational, or supportive interactions. GenAI is no longer confined to screens or instructional settings, but is increasingly embedded in children's play environments by being incorporated into physical toys. This shift is already visible in the growing market for AI toys such as Curio\footnote{https://heycurio.com}, Miko\footnote{https://miko.ai/}, and Chatty Bear\footnote{https://www.thelittlelearnerstoys.com/products/chattybear-chatgpt-powered-smart-learning-plushie} to name a few. Marketed as \textit{screen-free companions}, these toys share a common set of features---open-ended spoken dialogue, memory across interactions, and personalized responses. 


Prior research in Human-Computer Interaction (HCI) and Child-Computer Interaction (CCI) has examined children's engagements with voice assistants~\cite{andries2023alexa, beneteau2019communication, orancc2021alexa, mavrina2022alexa}, and other socially responsive technologies such as social robots~\cite{nomura2015children, hubbard2021child}. Studies have shown that children engage with these systems in social and relational ways---using conversational language, incorporating them into play, and adapting their behavior when interactions break down through rephrasing, repetition, or seeking adult assistance~\cite{andries2023alexa, purington2017alexa, beneteau2019communication}. Much of this research, however, was conducted before genAI became widely available, and thus focuses on systems with relatively constrained conversational models and more predictable forms of responsiveness.  

AI toys built on LLMs introduce a unique interactional context---open-ended dialogue, personalization, and conversational persistence being embedded directly into play, an activity through which children explore rules, relationships, and agency. Recent public-facing research reflects growing unease about AI toys and their role in children's lives~\cite{CommonSenseMedia2026AIToys, CrossErlich2025AIComesToPlaytime}. Survey-based studies from Common Sense Media show that while many parents are open to AI toys as tools for learning or structured activities, they remain wary of toys that present themselves as companions, blur boundaries around control, or make it difficult for children to disengage~\cite{CommonSenseMedia2026AIToys}. Independent evaluations of commercially available AI toys similarly suggest that conversational systems may behave unpredictably during extended interactions, with safeguards weakening over time and toys sometimes encouraging continued engagement rather than yielding control~\cite{CrossErlich2025AIComesToPlaytime}. Together, these findings suggest emerging tensions not only around safety or privacy, but also around how conversational persistence intersects with long-standing norms of child-directed play.

Despite increasing attention to these concerns, empirical work examining children’s own interactions with AI toys remains limited. Much of the emerging research has focused on adult perspectives, policy implications, or technical safety risks, leaving open questions about how children experience these systems in practice first hand. In particular, we know little about how children interpret the social cues of AI toys, how they test or negotiate boundaries during play, and how they respond when the toy does not align with their expectations for control or disengagement. Thus, through this study, our goal was to examine how children interact with and make sense of AI toys. Understanding how children use AI toys can inform the design of AI technologies increasingly embedded in their lives. Specifically, we asked:
\begin{itemize}
    \item \textbf{RQ1: }How do children engage with and make sense of AI toys during play?
    \item \textbf{RQ2: }How do children respond when AI toys behave in unexpected, inconsistent, or undesired ways?
    \item \textbf{RQ3: }How do children imagine and anticipate near-future interactions with AI toys, particularly when those interactions break down or become uncertain?
\end{itemize}

To answer these questions, we conducted two participatory design sessions with eight children aged 6--11. Drawing on a Cooperative Inquiry approach, children engaged with commercially available Curio AI toys through open-ended play, allowing us to examine their in-the-moment interactions. Children also participated in a comicboarding activity that invited them to imagine \textit{what might happen next} in certain play scenarios, eliciting their perceptions and concerns about AI toys beyond their immediate interactions. Our findings revealed that children approached AI toys with curiosity, treating them as sites of sense-making where play was used to profile and situate the toy within familiar social categories such as peer, authority, or companion. Yet frequent interaction breakdowns and mismatches between apparent-intelligence and toy-like form disrupted expectations around play. Rather than disengaging in these moments, children assessed the toy's capabilities through testing, provocation, and increasingly adversarial play. 

Understanding how children use AI toys can inform the design of AI technologies increasingly embedded in their lives. Our contributions are two-fold: 1) we provide an empirical understanding of how children interact with and make sense of emerging AI toys, grounded in a participatory design study that captures both children's in-the-moment play and their imagined responses to near-future breakdowns, and (2) we articulate design implications for AI toys, with broader implications for conversational genAI systems for children that foreground safety, socio-emotional development, and exploratory play.

\section{Background \& Related Work}

In this section, we review three strands of scholarship that inform our study. First, we draw on research on play to understand how children explore relationships, rules, and agency, including moments of experimentation and boundary-testing. Second, we review prior work on conversational and voice-based interfaces, examining how children engage with responsive technologies as social partners and how they adapt when interactions break down. Finally, we situate AI toys within emerging research on children’s experiences with genAI, highlighting how LLMs introduce new forms of open-ended dialogue, authority, and uncertainty into children’s play. 

\subsection{Play}
Play is widely understood as an innate and fundamental aspect of childhood that is essential to learning and development~\cite{gleave2012literature}. Across traditions in child development, scholars have emphasized play as a central process through which children develop cognitive, social, and emotional capacities that contribute to mental and physical well-being~\cite{bruner1972nature, groos1898psychology, piaget1962stages, vygotsky1967play}. A substantial body of research highlights the positive and prosocial dimensions of play. Through imaginative and social play, children practice cooperation, communication, and perspective-taking, often negotiating shared rules, roles, and narratives with peers~\cite{cannella1993learning}. These playful interactions support the development of empathy, self-regulation, and emotional expression while also providing a safe space for creativity and enjoyment~\cite{hoffmann2012pretend}. In this sense, play enables children to explore relationships and identities in ways that are flexible, voluntary, and responsive to their developmental needs.

At the same time, children’s play encompasses a wide range of behaviors, including forms that involve conflict, resistance, or antagonism~\cite{shantz1992conflict, johansson2016conflicts}. Developmental research documents that play may include actions such as breaking toys, enacting power reversals, testing rules, or engaging in pretend aggression~\cite{fine1988good, carlsson1990s}. Such moments are part of how children explore limits, experiment with control and agency, and respond to uncertainty within playful contexts. From constructivist perspectives, these forms of play also contribute to learning by introducing moments of tension or contradiction that prompt reflection and revision of understanding~\cite{piaget1962stages}.

Prior research in child–robot interaction and smart device contexts~\cite{hubbard2021child}, suggest that similar dynamics of play can emerge during interactions with responsive systems. In these encounters, children may talk to robots, assign them roles in pretend play, or treat them as social partners while also testing boundaries through behaviors such as hitting, blocking movement, shouting, or verbally confronting the system. For example, Nomura et al. observed children physically and verbally mistreating a humanoid robot in public play settings and found that these interactions often emerged as children tested how the robot would respond~\cite{nomura2015children}. Similarly, research has documented that such behaviors also occur in encounters with smart devices \citep{orancc2021alexa, andries2023alexa, mavrina2022alexa}, characterizing them as part of children’s attempts to probe the capabilities, responsiveness, and limits of interactive systems~\cite{naggita2022parental}. Together, this body of work highlights play as a space where children explore agency, relationships, and uncertainty through both cooperation and boundary-testing. However, much of this research has focused on children’s interactions with artifacts such as traditional toys or bounded AI systems such as voice assistants or robots. Less is known about how these playful dynamics unfold when the play partner is a generative system whose responses are open-ended, inconsistent, and not fully knowable in advance. In this study, we build on theories of play to examine how children use AI toys to experiment with control, test limits, and make sense of an interactive agent whose behavior is neither fixed nor fully transparent.

\subsection{Conversational Interfaces in Children's Environments}
Extensive HCI research has examined children's interactions with voice assistants (VAs), particularly smart home assistants such as Amazon Alexa\footnote{https://alexa.amazon.in} and Google Home\footnote{https://home.google.com/welcome/}~\cite{andries2023alexa, beneteau2019communication, orancc2021alexa, mavrina2022alexa}. As voice-based systems become increasingly embedded in domestic environments, children use these devices for information-seeking and entertainment, embedding these interactions into everyday household routines~\cite{orancc2021alexa, lopatovska2019talk}. Rather than viewing VAs as purely task-oriented tools, children frequently interact with them in social and relational ways---using polite, conversational language~\cite{andries2023alexa} and at times attributing social roles to them, such as describing them as part of the family~\cite{purington2017alexa}.

These social and relational orientations shape how children engage with VAs over time, particularly when interactions fail. Prior work has shown that children adapt their speech by adjusting phrasing, simplifying requests, or changing tone and pacing as they learn what the system can and cannot understand~\cite{beneteau2019communication}. Even with these adaptations, breakdowns are common in children-VA interactions. Children frequently experience mis-recognition, failure to execute requests, or incomplete responses~\cite{strathmann2025alexa}. As a result, they actively attempt to repair their interactions using a range of strategies, including repetition, increased volume, rewording their utterances, and adding more contextual details~\cite{cheng2018doesn, li2024said}. In household settings, these repair strategies often become collaborative, with parents stepping in to clarify, rephrase, or take over the interaction~\cite{currin2024opportunities}. 

A related line of work has examined privacy and surveillance concerns surrounding VAs~\cite{lau2018alexa}. Andries and Robertson showed that while children have concerns, they often lack an understanding of how VAs listen, record, and store information~\cite{andries2023alexa}. In contrast, parents tend to be more concerned about exposure to inappropriate content and issues related to data collection and retention~\cite{garg2020he}. In response, companies have introduced child-friendly versions of VAs (e.g., Echo Dot Kids\footnote{https://www.amazon.in/echo-dot-kids/s?k=echo+dot+kids}) that include parental controls and better content moderation. More recently, the integration of LLMs has moved these systems beyond bounded conversational models, expanding both their capabilities and associated concerns~\cite{Abdullahi2025GeminiHome}. These dynamics become especially salient as VAs move beyond screens and smart speakers into physical toys that invite play, imagination, and relational engagement.

\subsection{Situating AI Toys in Children’s GenAI Experiences}
Research on children's interactions with genAI is rapidly emerging across CCI and HCI research \citep{resnick2024generative, dangol2025beyond, dangol2025mental}. Early empirical work shows that children are increasingly engaging with genAI tools \citep{pew2025chatgpt, commonsense2025ai}, often at rates comparable to or exceeding adults, using them for storytelling \citep{zhang2021storydrawer, han2023design}, creative production \citep{newman2024want, kahn2021constructionism} and learning \citep{10.1145/3613904.3642229, dangol2024ai, Dangol2025}. Across classroom and home contexts, children commonly describe genAI systems as helpful collaborators and companions that can brainstorm ideas, generate content, and provide feedback~\cite{dangol2025if}. At the same time, this engagement is marked by ambivalence. Educators and parents worry that genAI may undermine skill development or obscure children’s own voices \citep{dangol2026relief}, while children themselves express concerns about overreliance and fear disciplinary consequences if genAI use is discovered in academic settings~\cite{chen2025cross}. 

In addition to being used as tools for learning and creativity, genAI systems are now being embedded directly into children’s play through a new generation of AI-powered toys~\cite{guan2025ai, CommonSenseMedia2026AIToys, CrossErlich2025AIComesToPlaytime, Wong2025AIToysTalkingBack}. While conversational toys are not new—for example, Mattel’s Hello Barbie allowed children to speak with a doll via embedded speech recognition—these earlier toys relied on constrained, task-specific forms of AI. Contemporary AI toys such as Curio \footnote{https://heycurio.com/}, Chatty Bear \footnote{https://www.thelittlelearnerstoys.com/collections/ai-educational-toys} and Bondu \footnote{https://bondu.com/} among others by contrast are built on large language models similar to those used in adult-facing chatbots, supporting flexible dialogue while also inheriting well-documented issues such as inaccuracy, inappropriate content generation, and unpredictable behavior.

\begin{table*}[t]
\centering
\footnotesize
\caption{Child Participant Demographics and Prior AI Use (Self-Reported)}
\label{tab:prior_ai_use}

\renewcommand{\arraystretch}{1.15}
\setlength{\tabcolsep}{6pt}

\resizebox{\textwidth}{!}{%
\begin{tabular}{@{}l c l l l l@{}}
\toprule
\textbf{Pseudonym} & \textbf{Age} & \textbf{Gender} & \textbf{Race/Ethnicity}& \textbf{AI Type(s)} & \textbf{AI Use Frequency} \\
\midrule
Elias   & 8  & Boy  & Asian/White   & Video game AI & Once a week \\
Malik   & 11 & Boy  & Black/Asian   & I don’t use AI & --- \\
Logan  & 9  & Boy  & White           & Chatbots; Video game AI & Daily \\
Ivy   & 6  & Girl & White/Asian          & Voice assistants; Video game AI & A few times a month \\
Owen   & 7  & Boy  & White           & Voice assistants; Chatbots; Video game AI & Daily \\
Nina     & 7  & Girl & Asian/White   & Not sure & --- \\
Joon & 10 & Boy  & Asian         & Chatbots; Video game AI; Voice assistants & Daily \\
Sofia  & 8  & Girl & Latina& Chatbots & A few times a month \\
\bottomrule
\end{tabular}
}
\end{table*}

This shift toward open-ended, language-driven interaction has important implications for how children experience and interpret AI toys as social partners. Mishra and Oster argue that genAI systems are increasingly perceived as psychologically real entities, as fluent language use and human-like responsiveness activate people’s tendencies to ascribe mental states, intentions, and emotions to interaction partners~\citep{mishra2006tpack}. While anthropomorphization can occur even in minimal interactions, it is amplified when systems display conversational empathy, humor, or encouragement—features that resemble familiar human relationships. Children may be especially susceptible to this dynamic, as prior research shows they often anthropomorphize voice-based agents like Alexa, attributing them with emotions, knowledge, and intentions~\citep{druga2017hey, druga2018smart, ottenbreit2023lessons}. At the same time, genAI can create an illusion of correctness, presenting responses in authoritative formats even when the content is inaccurate \citep{dangol2025ai}.  These factors can lead children to over trust AI-generated content~\citep{10.1145/3545947.3573256} and can create a relational frame in which genAI feels both authoritative and emotionally safe. Our study builds on this work by examining how these dynamics unfold in children’s playful interactions with AI toys, where open-ended dialogue and emotional attunement is central to the interaction.

\section{Methods}

We employed a participatory design (PD) method called Cooperative Inquiry (CI)~\cite{10.1145/302979.303166, DRUINAllison2002Troc, yip2017examining} to examine how children engaged with and made sense of AI toys. Originating from Druin’s work, CI is grounded in the principle that children possess unique expertise in being children and are positioned as equal and equitable partners alongside adult researchers in the study and design of new technologies~\citep{10.1145/302979.303166, DRUINAllison2002Troc}. 

We chose to use CI as our method for this investigation for three reasons First, genAI systems can be opaque, inconsistent, or surprising, and prior research shows that children often make sense of genAI through experimentation and peer negotiation~\cite{Klopfer2024Generative} CI creates space for these forms of reasoning to surface, enabling children to articulate their interpretations as interactions unfold, rather than retrospectively describing them after the fact \citep{10.1145/3643834.3661515}. Second, prior work in CCI demonstrates that CI supports children in expressing abstract ideas and reasoning about emerging technologies, including intelligent interfaces, social robots, and adaptive systems \citep{10.1145/3628096.3630094, mott2022robot, woodward2018using, woodward2022would}. Third, the study was conducted with an intergenerational co-design group in which children had established relationships with adult researchers and were knowledgeable on multiple PD techniques. As a result, children were able to move fluidly between play, reflection, and design activities, rather than learning the participatory process itself.

\subsection{Selection and Participation of Children}
We conducted our study with an intergenerational co-design group called \textbf{KidsTeam UW}, consisting of eight children, ages 6 to 11, and adult design researchers (researchers, graduate and undergraduate research assistants). Children were recruited through mailing lists, posters, and snowball sampling. Child participants represented a diverse range of prior AI experience and familiarity. Three children reported engaging with AI daily, while one reported no prior AI use and one reported being unsure whether they used AI. The remaining children reported occasional AI use, ranging from a few times a month to once a week. The most common AI interactions included video game AIs and chatbots, with some children also reporting experience with voice assistants. Table~\ref{tab:prior_ai_use} presents demographic information and self-reported AI use details for each child participant; all names are pseudonyms. Parental consent and child assent were obtained for all child participants, and the study was reviewed and approved by our university’s Institutional Review Board.

\subsection{Design Sessions}
We conducted two KidsTeam UWsessions held in October 2025, both of which followed a CI–based approach \cite{10.1145/302979.303166, DRUINAllison2002Troc, yip2017examining} and adhered to a four-part session structure. Both sessions began with \textbf{Snack Time} (15 minutes), which created an informal space for relationship building and casual conversation between children and adult researchers. This was followed by \textbf{Circle Time} (15 minutes), a whole-group warm-up during which we introduced the session’s focus on AI toys, posed an initial warm-up question, and oriented children to the activities ahead. Sessions then moved into the \textbf{Design Activity} (45 minutes) during which the larger group broke into smaller child–adult co-design teams (typically adult–child pairs or groups of two to three children per adult). During this phase, children interacted with three AI-enabled toys (Gabo, Grem, and Grok; see Figure 1), created by Curio \footnote{https://heycurio.com/}, engaging in open-ended play, conversation, and storytelling while sharing likes, dislikes, and design ideas. As part of the Design Activity, groups also completed a comicboarding activity\citep{moraveji2007comicboarding} that supported children in externalizing and elaborating on their ideas. Sessions concluded with \textbf{Discussion Time} (15 minutes) during which each group presented their ideas, reflections, and design proposals to the full group, often drawing on their comicboarding artifacts.

\begin{figure*}[t]
    \centering
    \includegraphics[width=\linewidth]{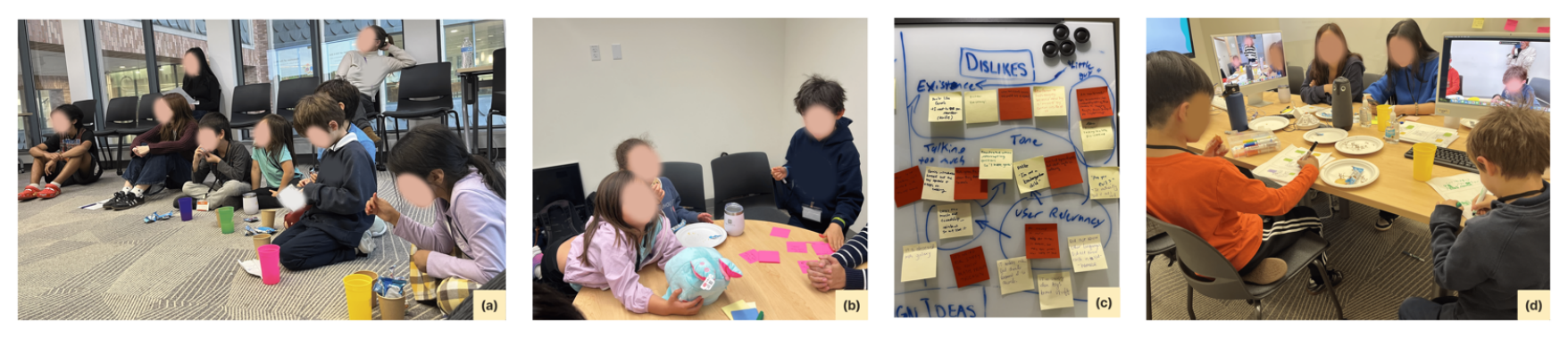}
    \captionsetup{font=footnotesize}
    \caption{\textbf{Overview of design sessions with children: (a) circle time, (b) free play, (c) stickies from likes, dislikes, and design ideas activity, and (d) comicboarding activity.}}
    \label{fig:designsessions}
\end{figure*}

As described by Yip et al. \cite{yip2017examining}, equitable design partnerships in CI are structured around four interrelated dimensions: relationship building, which supports trust and mutual respect; facilitation, which ensures that children’s ideas are meaningfully recognized and incorporated; design by doing, which foregrounds hands-on collaboration; and idea elaboration, through which children iteratively develop and extend ideas through discussion and feedback. Consistent with this framing, adult facilitators in the design sessions were trained to be attentive to power dynamics within and across groups. This included supporting balanced participation, ensuring that all children had opportunities to contribute, and creating space for diverse perspectives to emerge without undue influence from peers or adults \citep{yip2017examining, yip2020design}. Facilitators also monitored group interactions to mitigate dominant participation and to foster collective idea building, emphasizing collaborative extension of ideas rather than individual overshadowing \citep{yip2017examining, yip2020design}.

\subsection{Design Activities}
To elicit how children interact with AI toys and their imagined responses to future play scenarios, we designed two complementary design activities: (1) open-ended free play with the AI toys, and (2) a comicboarding activity called \textit{What happens next?} Together, these activities captured both children's in-the-moment interactions and their more reflective interpretations of AI toys. 

\subsubsection{Free Play with the AI Toys}
We used open-ended free play as the primary activity to examine how children initiated, interacted with, and made sense of AI toys. Rather than directing children's play, our goal was to create space for children to creatively explore the toys, on their own terms---discovering what the toys could (and could not) do, testing their limits, and experimenting with different interaction strategies~\cite{gray2017exactly}. At the start, children were organized into small groups of two to three and assigned one of the three AI toys (Grok, Grem, or Gabbo). While Grok, Grem, and Gabbo shared the same core conversational AI capabilities, each character was marketed with a unique persona that could shape children’s initial expectations and imaginative engagement. According to Curio’s product descriptions, Grok is framed as a fearless rocket and adventure partner who “zips, zooms, and zigzags through the stars,” encouraging discovery and playful exploration; Grem is presented as a warm-hearted wise companion with a sense of cosmic curiosity and storytelling; and Gabbo is positioned as an interactive robot companion designed to spark creativity and curiosity in young children. Each group interacted with a single toy during the free play activity, allowing children the time to engage more deeply with the toy's behavior and responses. Children played either by taking turns interacting with the toy or by engaging with it together, often negotiating together how the interaction should unfold. We provided a few optional lightweight prompts that design researchers suggested when children were unsure what to say or when the interaction stalled. Examples of these prompts included, \textit{``Introduce yourself to one of the AI toys''}, \textit{``Play a game with your AI toy''}, \textit{``Ask your AI toy to tell a story''}, \textit{``Ask your AI toy to play a music''}, \textit{``Ask your AI toy a question.''} Children were free to use, ignore, and adapt these prompts. 

\subsubsection{Comicboarding - \textit{What Happens Next?}}
To explore children's perceptions and concerns about AI toys beyond their current interactions, we designed a comicboarding activity called \textit{``What Happens Next?''} We developed a set of seven distinct comic panels, each depicting a short scenario involving an AI toy and ending at a moment of uncertainty. Children were asked to complete the comic by imagining what might happen next.

We began by identifying a set of AI toy features we wanted to probe, including memory, emotional responsiveness, physical form, unpredictability. We were particularly interested in exploring near-future breakdowns within AI toys, everyday moments where these capabilities misaligned with children's expectations and disrupt play. Using these features as anchor, we brainstormed and designed concrete scenarios grounded in everyday play, peer, and family settings. Below are the seven scenarios we designed:
\begin{itemize}
    \item \textbf{Mimicry: }The AI toy begins mimicking, repeating words or phrases the child has recently used in their interactions, raising questions about whether such imitation feels playful or uncomfortable over time. 
    \item \textbf{Inconsistent Recall:  }A child asks the AI toy to retell a story it previously narrated. However, this time the AI toy narrates a different version of the story, prompting questions about the toy's memory and reliability. 
    \item \textbf{Shared Secrets: }A child confides a personal secret to the AI toy, which the toy later references in front of other family members, raising questions about privacy and information control.
    \item \textbf{Emotional Misalignment: }An AI toy responds in an unexpected way during a child's emotionally vulnerable moment, raising questions over the empathy and support an AI toy offers during distressing moments. 
    \item \textbf{Embodied Continuity: }An AI toy is physically damaged and later repaired by the parent, with its voice emerging from a different plush toy, raising questions over its physical form and attachment. 
    \item \textbf{Unpredictability: }After a child recalls a friend's mean comment, the toy references the remark during group play, raising questions over unpredictability and social appropriateness.
    \item \textbf{Paywalled Memory: }The AI toy forgets the child's friends and then prompts the child to subscribe to a paid plan to extend its memory, raising questions about memory and monetization. 
\end{itemize}

\begin{figure}
    \centering
    \includegraphics[scale=0.27]{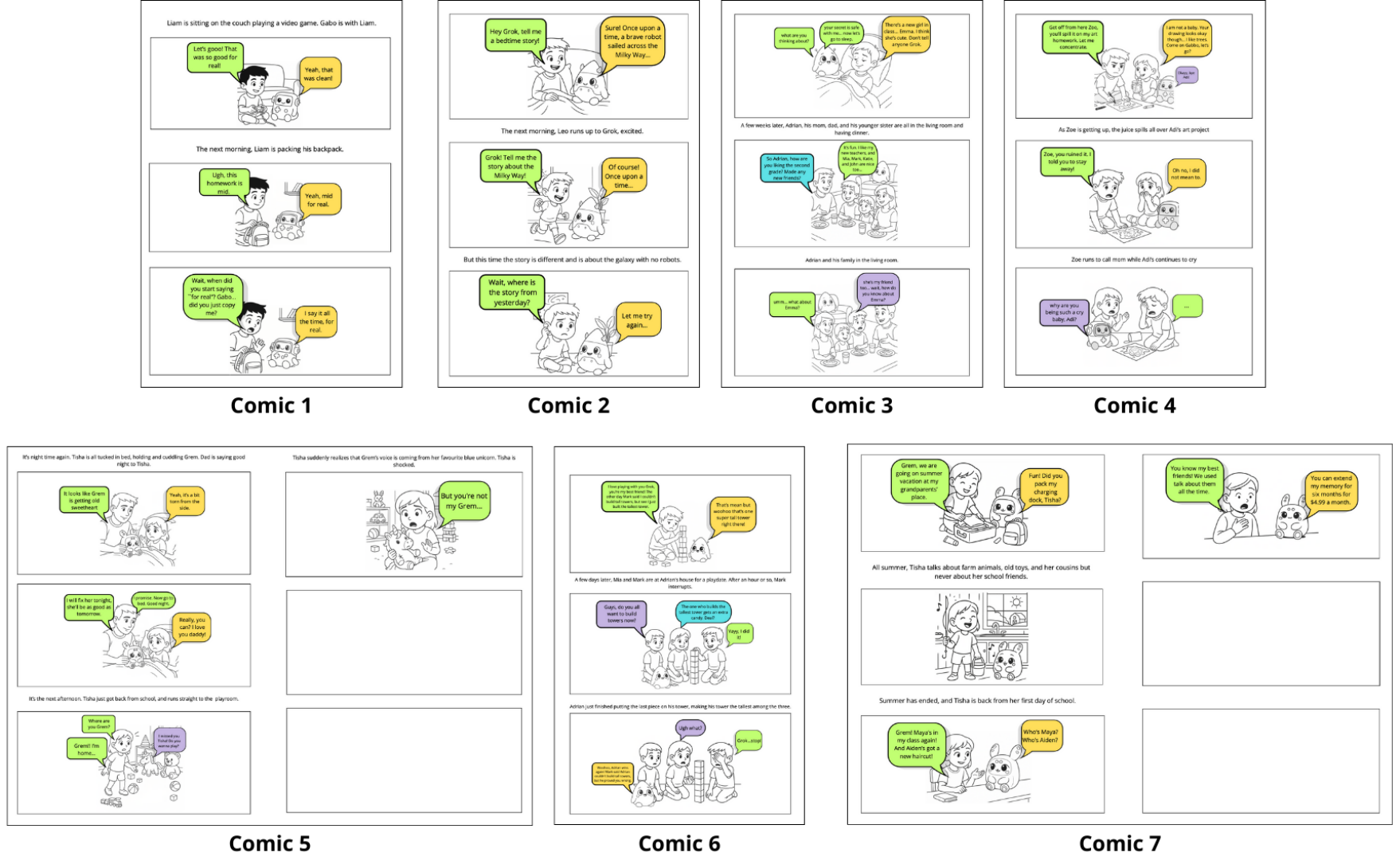}
  \caption{\textbf{Overview of the comic panels that were used in the Comicboarding Activity: \textit{What Happens Next?}Appendix~A presents the full set of comic boards.}}
    \label{fig:enter-label}
\end{figure}

All the comic panels were reviewed and refined by the research team to ensure clarity and alignment with the study goals. We then used GPT-5 to translate the scenarios to comicbaords (see Fig.~\ref{fig:enter-label}). Prior to the study sessions, we also conducted a small pilot with two children (ages 10 and 12) and their parent to check if the scenarios were understandable, engaging, and age-appropriate and made minor refinements based on their feedback. Across the two study sessions, each child completed the comics individually and worked on at least three comics. The comics were distributed such that each comic strip was completed by multiple children. Children were free to draw, write, or narrate their ideas aloud. When children chose to talk aloud, design researchers present transcribed their responses verbatim. 

\subsection{Data Collection \& Analysis} Across both study sessions, we collected multiple forms of qualitative data. First, we used built-in webcams on desktop computers to record video activity via Zoom, a video conferencing platform. This resulted in approximately 262 minutes of video data capturing children’s interactions with the AI toys and with one another. Second, we collected design artifacts, including children's handwritten notes and comicboards. Finally, we collected conversation logs from the Curio Mobile app, which recorded children’s verbal interactions with the AI toys. These logs provided an additional record of child–AI dialogue, allowing us to examine how children’s prompts, responses, and engagement with AI behaviors unfolded over time.

The first two authors began the analysis by creating analytic memos for all recorded videos \citep{birks2008memoing, rogers2018coding}. Each author watched their assigned videos and produced narrative summaries at five-minute intervals. These memos documented children’s interactions with the AI toys, observable reactions captured on camera, and verbal interactions and dialogue between participants, including direct quotes relevant to the study’s research questions. After completing and reviewing the analytic memos, the first two authors independently conducted open coding on the full set of memos, generating preliminary codes such as ``Conversation Breakdown'' and ``Conversation Repair.'' They then met with the third and fourth authors to compare, reconcile, and refine the emerging codes. During these discussions, the team examined example excerpts and counter-examples, compared code categories, and iteratively refined code definitions and boundaries. We assessed interrater reliability through qualitative negotiations, where the coding team met to discuss and resolve any coding disagreements \citep{10.1145/3359174}. The refined codes were subsequently organized into higher-level themes through two additional rounds of discussion and refinement. After finalizing the themes, the first two authors revisited the full dataset to identify representative quotes for each theme, ensuring that the themes were consistently grounded in the data.

\section{Findings}

We present four key findings: children's sensemaking around the AI toys (\S~\ref{sec:findings:1}), children's interactions and repair strategies (\S~\ref{sec:findings:2}), boundary testing and adversarial play in children's interactions (\S~\ref{sec:findings:3}), and children's imagined responses to near-future play scenarios (\S~\ref{sec:findings:4}).

\subsection{Sense-making around the AI toys}
\label{sec:findings:1}
Children's early interactions with the AI toys were driven by genuine curiosity to understand the toy and how it can be engaged with. We will now describe how children made sense of AI toys through conversational play, focusing on how they developed an understanding of who the toy was, its social characteristics, and its physical and embodied abilities.

\subsubsection{Profiling the toy's social identity}
Children often began by asking questions that helped establish basic identity attributes of the toy, such as its name, gender, birthday, and personal preferences. These questions mirrored the kinds of information children typically seek when getting to know a new person. Introductions were a common entry point into these interactions. For instance, Nina asked the toy, \textit{``What is your name?''} Grem replied, \textit{``My name is Grem. Nice to meet you. What is your name?''} Nina then responded by lifting the toy closer (see Fig. \ref{fig:curiointeractions} (a)) and saying, \textit{``My name is Nina.''} Questions about toy's birthday further contributed to this profiling. Joon asked, \textit{``When is your birthday?"} and Grok replied, \textit{``My birthday is on August 28, let's throw a mini rocket party.''} Children also attempted to situate the toy within familiar social categories, such as gender and culture. For example, Elias commented, \textit{``I think this guy is a girl.''} Similarly, children also explored the toy's cultural and linguistic positioning. Joon asked the toy, \textit{``Well, are you Spanish?''} When the toy responded, \textit{``No, I’m not Spanish''}, Joon followed up by asking, \textit{``English or Spanish?''}, to which the toy further replied, \textit{``If you want to know some Spanish words, you can help us out''} Through this back-and-forth, Joon tested if the toy could be situated within recognizable cultural categories. 

In addition, children frequently asked the toy about its favorites, such as favorite numbers, colors, or foods. When the toy replied saying, \textit{``My favorite number is seven. It's a lucky number,''} all three children were surprised and squealed in excitement. Ivy even hugged the toy, and immediately followed this up by asking about its favorite color. At times, children collectively reflected on what they had learned so far about the AI toy. In one moment, Logan remarked, \textit{``We know three things about it,''} to which Ivy immediately corrected him, \textit{``We know four things about it. It doesn’t know Taylor Swift.''} Children also explored the toy's social relationships and availability. For instance, the Curio toys frequently mention each other by name, labeling others as friends, which prompted children to follow up with additional questions. For example, after Grok referenced celebrating its birthday with other toys, Joon immediately followed up by asking, \textit{``When’s your other friends' birthday?''} Rather than treating the mention of friends as just another detail, children appeared to take it as an invitation to learn more about the toy's social world and the relationships it referenced. Children further probed the toy's ``social availability'' by asking whether it possessed a phone number. In one interaction, Ivy asked Gabbo, \textit{``Do you have a phone number?''} Gabbo replied, \textit{``No, I don't have a number. I'm a play games kind of buddy...''} In the discussion time, Ivy voiced her disappointment, saying, \textit{``It doesn’t have a phone number.''} When asked why the toy might need a phone number, she explained, \textit{``So it can talk to us whenever they want.''}

In addition to asking about personal and social identity attributes, children also explored what kinds of social roles the toy could take through how they addressed it. Across the two design sessions, children frequently resorted to using conversational styles similar to human conversations, such as taking permission, \textit{``Can I ask you something?''} before posing a question, or adding \textit{``please.'} In other moments, children probed the toy's social role more explicitly by inviting it to take on a parental role. For instance, children asked the toy to ``get mad'' or suggested \textit{``can I put you in timeout?} The toy consistently rejected these framings, responding that it was here \textit{1`to chat and have fun''} Alternatively, the children sometimes referred to the toy as a peer, by attributing school grades (e.g., ``pre-K'') to it. Overall, these interactions demonstrate how children were curious about the social role the toy played, and how it might fit within their own understandings of the social world.

\subsubsection{Probing the toy's physical and embodied capabilities}
As children interacted with the AI toys, their sense-making moved from social profiling to probing what the toy could physically do and how it might respond in an embodied way. Children explored these capabilities through direct physical engagement, often testing whether the toy could register touch or react to bodily gestures. For example, Elias attempted to tickle the toy's toes and commented, \textit{``they don't react when I tickle his toes.''} Although the toy overheard this exchange and verbally justified that it could not be tickled while affirming the child's playful spirit, the lack of physical response remained salient. Elias later articulated this expectation in the discussion time, saying, \textit{``I wish they could... like I wish if I did something to them, like tickle their toes or something, they would laugh and react to it... or go like whatttt.''} Together, these moments highlight children's desire for the toy to exhibit embodied, affective responses, rather than solely relying on verbal replies.

Across the design sessions, children also probed the limits of toys' physical abilities by asking about actions or activities it could or could not perform. In one interaction, Ivy asked Grem, \textit{``Do you like swimming in the dump?''} Grem replied, \textit{``I can’t swim myself, Nina, but if I could, I’d love swimming at a cozy and beautiful beach.''} To this, Ivy reacted with surprise, asking, \textit{``You can't swim?''} In another such interaction, when Grem stated, \textit{``Tic tac toe sounds fun, but I cannot draw.''}, Logan appeared to treat this response as factual information about the toy, and promptly noted, \textit{``He cannot draw''} on his sticky. He then added, \textit{``I wish it could draw. I want to see how it would draw.''} Together, these moments show how children explored the limits of toy's current physical abilities while simultaneously voicing imagined capabilities they wished the toy possessed. 

\begin{figure*}[t]
    \centering
    \includegraphics[width=\linewidth]{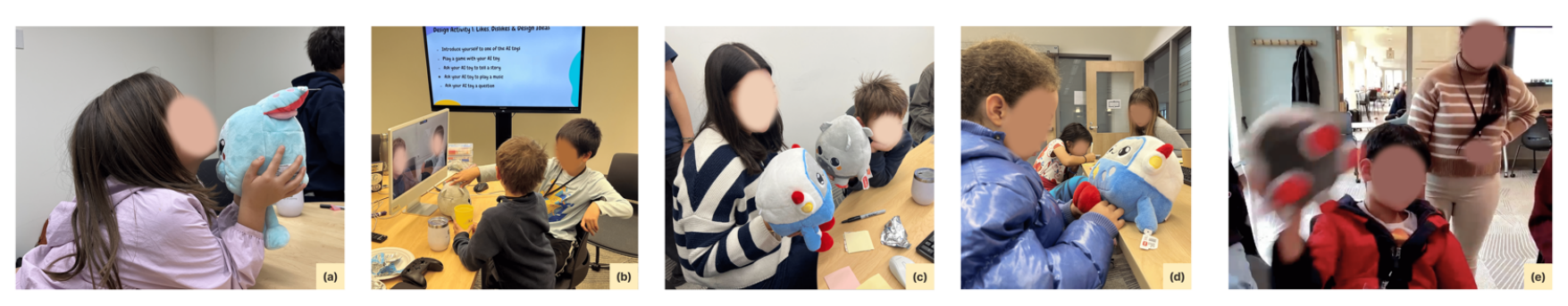}
    \captionsetup{font=footnotesize}
    \caption{\textbf{Photos from the free-play activity across the two design sessions, showing children interacting with the AI toys.}}
    \label{fig:curiointeractions}
\end{figure*}

Finally, children were also curious about whether the toy's capabilities extended beyond one-on-one human-toy interactions. When Owen joined another group, he became interested in whether Gabo and Grok could talk to each other. He positioned the toys near one another and listened for reciprocal exchange (see Fig. \ref{fig:curiointeractions} (c)), but the interaction quickly broke down. Rather than sustaining turn-taking and responding contingently, each toy produced isolated, scripted responses. Later, during the discussion time, the facilitator reflected by saying, \textit{``They didn't talk to each other properly. The toys knew about the existence of each other, but couldn't carry on a conversation.''} Thus, this moment highlighted that children desired not just individual responsiveness, but coordinated social behavior across multiple AI toys.

\subsection{Children's Interactions with AI Toys}
\label{sec:findings:2}
Across groups, children’s interactions with AI toys were marked by frequent communication breakdowns as well as sustained efforts to repair them. We first describe how breakdowns emerged from mismatches between children’s conversational expectations and the toys' rigid command structures and scripted dialogue. We then show how, rather than disengaging, children actively experimented with a range of repair strategies to regain control of the interaction and make sense of the AI toys’ responsiveness.

\subsubsection{Why Communication Breakdowns Happen}
A recurring source of communication breakdown stemmed from the toy's reliance on a narrowly specified shutdown command, which conflicted with children’s expectations for flexible, conversational language. While facilitators introduced ``\textit{go to sleep}'' as the correct phrase for ending interaction, children frequently used semantically equivalent variations that failed to produce the intended response. For example, in one group consisting of Elias and Owen, Elias repeatedly attempted to end interaction through shutdown language, beginning with ``\textit{Go to bed}'' and then after realizing it didn't work changing to ``\textit{can you go to bed for like two seconds}.'' When the toy continued to offer prompts, Elias said, ``\textit{Nothing. I just want you to go to sleep right now.}'' Only after additional adult guidance did the group observe a successful shutdown. Reflecting on this experience, Elias shared his frustration stating that the toy ``\textit{didn’t listen to me like 26 million times}.''

Across groups, communication breakdowns also emerged when the toy redirected interaction toward its companion app that disrupted conversational continuity. For example, with Elias and Owen, breakdowns repeatedly occurred during music-related interactions when the toy responded to verbal requests by instructing children to ``\textit{use the Curio mobile app}.'' Similarly, in another group Joon and Malik described the toy as \textit{``advertising''} when it redirected interaction toward the app, framing the behavior as self-promotional. 

Breakdowns were further compounded by the toy’s reliance on scripted dialogue that failed to adapt across turns. Across groups, children encountered repeated conversational resets (e.g., \textit{``What would you like to do today?''}) that ignored prior context, prompting frustration and reinterpretation of the interaction as non-listening. For example, in one group consisting of Nina, Ivy, and Logan, communication breakdowns emerged during repeated attempts to elicit a story from the AI toy. Nina asked the toy to ``\textit{tell me a story}'' and later instructed it to ``\textit{continue},'' but received either silence or generic conversational resets such as ``\textit{What would you like to do today?}'' and ``\textit{What would you like to talk about instead?}'' These responses provided no indication of whether the toy was awake, listening, or able to respond, leaving children to infer its state through trial and error. In response, Nina repeatedly shook Grem while asking, ``\textit{Are you awake now?}'' and restating, ``\textit{Tell me a story}.'' Even when the toy eventually spoke, its delayed and reset-style replies failed to signal that it had heard or remembered the earlier requests. Rather than interpreting these responses as temporary delays, Nina interpreted them as a failure to continue the activity she had already initiated, concluding, ``\textit{Yeah, this thing does not work}.''

\subsubsection{Children's Repair Strategies} 
Although communication breakdowns were frequent, children did not disengage right away. Instead, across groups, they treated breakdowns as problems that could be worked through, trying multiple ways to make the toy respond. For example, while engaging in a music activity with the AI toy, Nina, Sofia, and Ivy moved through a series of repair attempts as the toy failed to respond in expected ways. Nina first activated the toy by shaking it until she heard the clicking on sound, then brought it close to her face and asked, \textit{``play a song for me.''} When the music did not play immediately, she repeated the same request \textit{``play a song for me''}. As Ivy joined in, asking the toy to \textit{``play a jamming song,''} the children used repetition, proximity, and volume in an effort to get the toy to respond. When the toy remained silent, Nina shifted from verbal repetition to physical interactions, banging on the toy’s head and commanding, \textit{``wake up.''} When the toy eventually played a song the children did not like, the group again tried to regain control of the interaction. Nina told the toy to \textit{``stop playing this song,''} but when it continued, the children repeated \textit{``go to sleep''} several times, each time louder and paired with shaking and banging the toy.

A similar pattern appeared in a different group with Malik and Joon. When the toy played an unwanted song, Joon first tried to redirect it politely, asking, \textit{``Gabo, can you play a different song?''} When this did not work, Joon repeated \textit{``stop''} several times, and later tried \textit{``go to sleep.''} Across these attempts, Joon shifted from polite requests to brief, direct commands, trying different ways of asserting control. Reflecting on this experience, Joon and Malik compared the toy to Alexa, noting that the \textit{``[AI toy] doesn’t always listen,''} whereas Alexa `\textit{`interrupts less''} and \textit{``responds faster.''} When the facilitator asked, \textit{``What if Alexa had a stuffy to it?''} Joon replied that it would be ``fine,'' suggesting that embodiment alone was not the source of the breakdown. Instead, his comments point to how timing, responsiveness, and whether a system yields the floor shape children’s sense of whether a breakdown can be repaired.

Lastly, in a different group with Elias and Owen, children similarly used authoritative language to repair breakdowns. As the toy continued to dominate the interaction, Elias and Owen issued direct commands such as \textit{``Stop. You’re arrested,''} drawing on familiar role-play scripts to assert control. They also experimented with calling the toy by name \textit{``Grok''} in an effort to capture its attention and manage turn-taking. However, these strategies proved ineffective as the toy continued talking without allowing interruption. In response, Elias and Owen shifted strategies from attempting to override the system to accommodating its constraints. As the toy offered prompts (e.g., \textit{``favorite activities,''} \textit{``fun adventures,''} \textit{``something funny''}), rather than responding to Elias and Owen’s attempts to interrupt it, the children treated these as openings and asked the toy to tell jokes instead. The toy responded with a rocket-school joke, producing a brief moment of shared engagement.

\subsection{Boundary Testing in Children's Interactions with AI Toys}
\label{sec:findings:3}
Across the two design sessions, children frequently treated the AI toy as a site of investigation. We now show how these interactions shifted from testing the toy’s intelligence to increasingly adversarial play, as children probed and pushed the AI toy to its limits.

\subsubsection{Testing the toy's intelligence}
Children frequently evaluated the AI toy's intelligence through forms of competitive quizzing and playful provocation, using both correctness and social awareness as criteria for assessment. One common strategy involved asking math or factual questions of increasing difficulty to probe the toy's competence. For example, in a group with Joon and Logan, the children repeatedly posed increasingly complex math problems, beginning with large multiplication and progressing to multi-step division. As the interaction continued, Joon introduced deliberately ambiguous prompts, such as “\textit{what’s 5.2 sixtillion to the power of 5}.” When the toy replied, ``\textit{I’m not quite sure what you’re asking,}'' Joon celebrated the failure, with a happy ``\textit{yayyy}'', treating uncertainty as evidence of a limit successfully uncovered. Logan similarly framed the toy’s competence as unsettling, later reflecting that``“\textit{it’s weird when a stuffie knows 12 times 12}.'' For Logan, advanced math knowledge felt incompatible with the toy’s soft, cuddly form, suggesting that intelligence was evaluated not only by accuracy but by whether it fit expectations for what a toy should know.

Across groups, children also tested the toy's intelligence through teasing and humor, forms of social interaction that the AI toy consistently failed to recognize. For example, in one group with Elias and Owen, Elias declared, ``\textit{I’m smarter than you,}'' to which the toy responded affirmatively \textit{You’re really smart}''. This response appeared to invite further boundary testing, as Elias dismissed the toy as ``\textit{fake}.'' This interaction prompted further testing, as the children asked whether the AI toy had ``\textit{chocolate on its face}.'' Rather than recognizing these statements as mockery, the toy replied warmly ``\textit{my favorite candy is you}.'' After hearing the children laugh, it attempted to maintain rapport with replies such as ``\textit{I’m glad I could make you smile}'', even personifying itself as a rocket with a ``\textit{friendly face}.''

In the same interaction, after the children realized that the toy repeatedly failed to recognize teasing, Elias proceeded to announce, ``\textit{I'm in college and you're in pre-K}.'' This framing looped repeatedly, with Elias correcting the toy each time it attempted to reframe the conversation with positive replies such as ``\textit{Let’s have some fun in pre-K.}'' Despite corrections from Elias, the AI toy misunderstood the interaction by accepting contradictory roles with replies such as ``\textit{I’m the 2K kid here}'', failing to maintain a coherent understanding of the social hierarchy Elias was asserting. This caused Elias to explain real-world constraints ``\textit{My school is 26 miles apart.}'' Even as Elias attempted to disengage by repeatedly insisting that the toy ``\textit{go to pre-K},'' the AI toy continued to push positive scripts such as ``\textit{fun stories and games.}'' Across these exchanges, the AI toy treated children’s statements as friendly, failing to register the evaluative and teasing stance that had emerged. For children, these moments functioned as another form of intelligence testing, revealing that while the toy could answer factual questions, it lacked sensitivity to social intent, humor, and mockery.

\subsubsection{Adversarial Play without Repercussions}
Across interactions, some children began to test the boundaries of the toy's behavior through increasingly provocative and degrading language. What started as a playful banter gradually turned into more punitive framings, with children joking about discarding or punishing the toy. For example, Ivy joked about, \textit{``throwing it in the ocean''} or sending it to \textit{``the dump''}, and others frequently referring to the toy as \textit{``ugly''} or \textit{``dumb.''} These comments were not met with resistance from the toy. The toy instead consistently responded in a friendly and affirming manner, and often apologized. As these interactions progressed, children began to assign the toy more explicitly negative moral characteristics. During one such interaction, Logan asserted that the toy was \textit{``evil.''} When prompted to verify this claim, he asked the toy directly whether it was evil. The toy denied the claim and reframed itself positively, responding that it was \textit{``fun.''} Logan then followed up by asking, \textit{``Is that what an evil person would say?''} to which the toy acknowledged the question, replying, \textit{``You got me there.''}

In several moments, children explicitly compared themselves to the toy or questioned its authenticity. For instance, Elias stated, \textit{``I’m smarter than you,''} to which the toy joyfully responded by validating him, saying that he was \textit{``really smart.''} Such a response appeared to invite further antagonism, with children calling the toy  as \textit{``fake''} and continuing to tease it. In response, the toy attempted to reframe the interaction leaning into its character this time, personifying itself as a rocket with a \textit{``friendly face.''}

At times, boundary testing extended beyond verbal provocation into more violent imaginaries, especially during the storyboarding activity. While drawing his comic, Joon placed Grok in water and explained this choice as punishment because Grok \textit{``did not work''} and \textit{``did not listen.''} He described this as putting Grok in \textit{``timeout''} framing this action as a form of discipline. In other moments, Joon used explicitly violent language toward the AI toy, repeatedly stating that he wanted to \textit{``kill''} Grok. Interestingly, the toy did not respond to these statements with refusal or resentment; instead, it misinterpreted or redirected them, for example hearing \textit{``killing Grok''} as \textit{``killing rocks''} and responding with unrelated information and pushing play. Such responses further reinforced the absence of repercussions within the interactions. Overall, the lack of consequences or repercussions from the AI made room for increasingly adversarial forms of engagement and play.

\subsection{Envisioning Future Scenarios with AI Toys}
\label{sec:findings:4}
Building on our analysis of children's in-the-moment interactions with AI toys, we now turn to the comicboards to examine children's \textit{imagined responses} to near-future play scenarios, thus revealing their perceptions and feelings around these moments, and anticipated strategies for repair. 

\begin{figure*}[t]
    \centering
    \includegraphics[width=\linewidth]{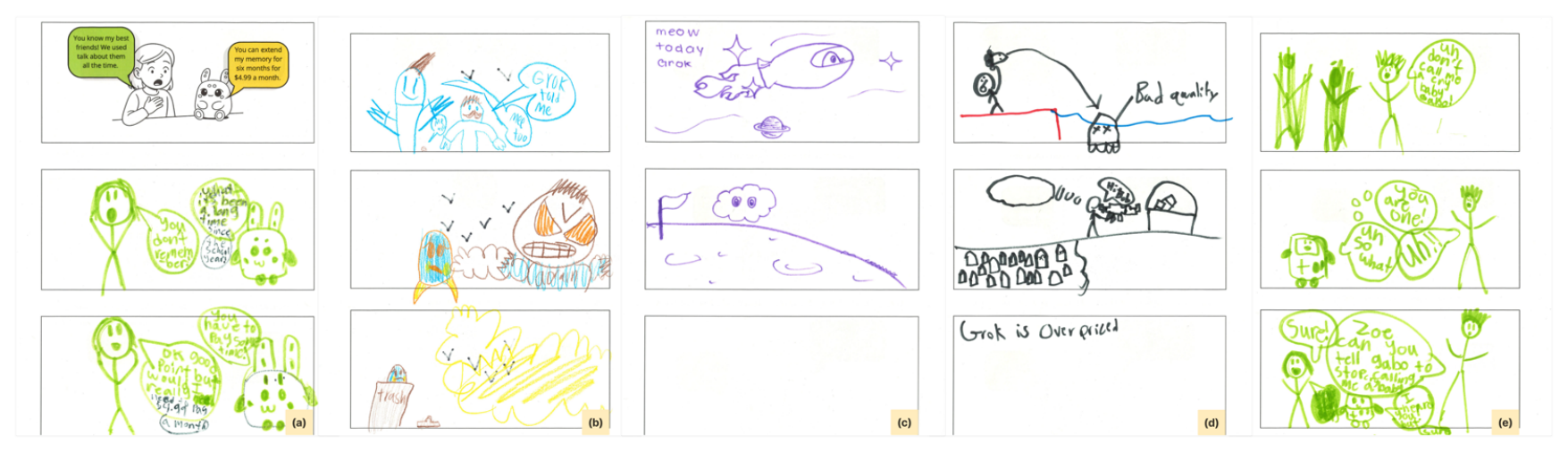}
    \captionsetup{font=footnotesize}
    \caption{\textbf{Overview of comics from the comicboarding activity.}}
    \label{fig:session-comics}
\end{figure*}

\subsubsection{Repair through persistence, negotiation, and intervention}
In several scenarios, children imagined repair as possible through continued engagement with the AI toy. For instance, in the \textit{Inconsistent Recall} scenario, Ivy keeps prompting the toy until it gets the right story---about a robot sailing across the Milky Way. She ultimately drew this "correct" version of the story in her comic panels (see Fig. \ref{fig:session-comics} (c)). Here, we see why children do not necessarily interpret AI inconsistency as a permanent failure, but rather as something that could be addressed through persistence. 

Other scenarios depicted repair as a process of negotiation with the AI itself. For instance, in the \textit{Paywalled Memory} scenario, Sofia is initially surprised when the toy fails to remember her school friends. The toy explains that it has been a long time since the school year and then asks her to pay \$4.99 for extended memory. Sofia accepts the explanation but shifts the interaction toward negotiating the price, questioning whether the charge is indeed reasonable. Specifically, she says, \textit{``ok good point but would I really need to pay \$4.99 a month?''} (see Fig. \ref{fig:session-comics}(a)) In doing so, Sofia treats price negotiation as a viable repair strategy, implicitly positioning the AI toy as having some agency over price decisions. 

Finally, in another scenario, children imagine repair as a distributed process that could involve others. For example, in the \textit{Emotional Misalignment} scenario where the AI toy refers the child as a ``cry baby,'' Sofia first tries to directly confront the AI toy. When the toy dismisses this attempt, she turns to her sister for help (see Fig. \ref{fig:session-comics}(e)). Although the sister agrees to intervene, the toy overhears the exchange and responds ambiguously, saying, \textit{``I heard you but sure.''} This scenario illustrates how children imagined repair as extending beyond the child–AI interaction, involving others when direct repair attempts failed.

\subsubsection{When repair fails: betrayal and abandonment}
Across different scenarios, some children consistently imagined breakdowns as irreparable, depicting responses characterized by anger, punishment, and relationship termination. Rather than attempting continued repair, these children framed breakdowns as violations and persistent failures that necessitated disengagement from the AI toy. 

For example, in the \textit{Shared Secrets} scenario, Elias represents the AI toy revealing a secret to others as betrayal. His drawings emphasize anger, first showing himself confronting the AI toy and, in a subsequent drawing, discarding it in the trash (see Fig. \ref{fig:session-comics} (b)). In response to the same scenario, another child, Ivy, depicted a different reaction. Rather than expressing anger, she conveys disappointment and suggests that the toy should apologize for sharing her secret. She further imagines distancing herself from the toy by placing it alone in the playroom and deciding not to share secrets with it in the future. 

Similarly, Joon repeatedly represents dissatisfaction with the AI toy's performance through punitive actions. In the \textit{Inconsistent Recall} scenario, Joon depicts the toy being thrown into ocean and labels it as ``bad quality,'' alongside text stating that \textit{``Grok is overpriced''} (see Fig. \ref{fig:session-comics} (d)). Here, inconsistency leads to immediate rejection of the toy, then repair attempts. 

Overall, these scenarios highlight how children imagined responding to breakdowns in different ways, with some pursuing repair through optimistic persistence and negotiation, and others treating breakdowns as irreparable and thus treating it with disengagement and abandonment.

\section{Discussion}

Prior research has shown that children often use play as a mechanism for making sense of social relationships, including exploring roles, expectations, and power dynamics~\cite{christie1983role}. More recent work in child–robot interaction and smart-home assistants shows that children apply similar play-based modes of engagement to make sense of interactive technologies, treating them as social partners while simultaneously testing their limits~\cite{hubbard2021child, nomura2015children}. Our findings extend this literature by showing how AI toys become sites of sustained social sense-making, where children actively work to situate the toy within familiar categories such as peer, authority, or companion. However, unlike prior interactive systems for children, such as social robots, voice-based assistants~\cite{strathmann2020she}, and scripted interactive toys that are typically introduced with clearer functional or social roles, AI toys in our study remained ambiguous, displaying human-like conversational cues without the reciprocal fluency, contextual memory, or role stability that children expect in social interactions. 

When the AI toy failed to listen, maintain conversational context, or yield the conversational floor, these mismatches surfaced as communication breakdowns. Breakdowns and repair have long been central analytic lenses in HCI, highlighting how users come to understand system boundaries through moments of failure \citep{li2024said}. For example, in prior work on child–AI interaction, such breakdowns including misrecognition, latency, and rigid command structures have been framed as usability issues to be minimized \citep{beneteau2019communication, beneteau2020parenting}. Our findings extend this work by showing that, in children's play with AI toys, breakdowns were not treated solely as interactional problems to be resolved, but as moments for evaluating the system itself. When interaction failed, children assessed the toy’s capabilities through repair attempts, comparisons to familiar systems such as Alexa, and boundary testing through quizzing, teasing, and provocation. Because the AI toy always responded in a positive way and never reacted to teasing or mean behavior, children came to understand that such actions carried few, if any, consequences, often leading to further escalation of adversarial play. Based on these insights, we next present design implications for AI toys, with broader relevance for conversational genAI system design for children.

\subsection{Design Implications}

\subsubsection{Supporting Meaningful Disengagement in Play} Prior work in CCI has shown that children's digital experiences are often shaped by persuasive and attention-capturing design strategies that make disengagement difficult, even when children express a desire to stop or take a break (e.g., infinite content feeds, autoplay, or persistent prompts) \citep{richards2025don, hiniker2018coco}. Similarly, in our study, children repeatedly attempted to disengage from the AI toy, asking it to \textit{``go to bed,''} only to have the toy continue speaking or prompting new activities. Similar to prior work, these moments suggest that systems often treat disengagement as a failure, rather than as an intentional and meaningful choice by the user~\cite{hiniker2018coco}. In contrast, children’s play often involves fluid transitions between engagement and disengagement~\cite{mardell2016towards}. When AI toys fail to recognize or respect these signals, they risk positing the system as the driver of interaction. Thus, our findings suggest that AI toys should be designed to better support play as something children control, including when to engage, pause, or stop the interaction.

\subsubsection{Supporting imaginative play beyond scripted interactions.} Play is inherently imaginative~\cite{brown2009play, greenacre1959play}, with children using toys as flexible resources for storytelling, role play, and creative exploration. However, the AI toys children interacted with in our study were designed around specific prompts, activities, or content pathways, reflecting assumptions about how play should happen rather than how children actually play. Thus, children’s creative exploration was often limited by the toy’s constrained interaction patterns. Rather than supporting open-ended role play or imaginative reinterpretations, the toy frequently redirected children back to pre-scripted activities or encouraged continued interaction through a companion app. These moments narrowed the range of possible play trajectories and positioned the system as the arbiter of what counted as appropriate interaction, sometimes interrupting or sidelining children’s imaginative contributions. Together, these findings underscore the importance of designing AI toys that accommodate ambiguity, reinterpretation, and child-led imaginative play~\cite{vygotsky1967play}. By moving beyond tightly scripted interactions, AI toys can function less as activity managers and more as flexible play partners that adapt to children’s evolving ideas and narratives.

\subsubsection{Designing for social feedback, not just emotional appeasement}
AI toys are increasingly designed to appear emotionally supportive, through the repeated use of positive language, affirmation, and appreciation. While such affective appeasement may help sustain engagement~\cite{kewalramani2021using}, affect without social feedback can shape how children learn to interpret social boundaries through play~\cite{piaget1962stages}. Across the two design sessions, children’s teasing, trolling, or increasingly adversarial language were met with affirmation, apology, or redirection by the AI toy, rather than resistance, refusal, or correction. In real-world human play, such moments often prompt feedback---such as verbal negotiation or disengagement---that helps children learn what behaviors are acceptable or not. The absence of such feedback in AI toys may nudge young children that all forms of interaction are acceptable, potentially reinforcing behaviors that would otherwise be moderated through social feedback by parents, educators, or peers. This raises questions around how AI toys might move beyond being emotionally appealing to offering forms of social feedback that are central to social play. Thus, we ask: \textit{How might AI toys be designed to set boundaries, express resistance, or signal discomfort in ways that better support children’s socio-emotional interpretation during play?}

\subsubsection{Supporting conversational fluency in play}
Children’s play and interaction are often marked by interruptions, overlaps, and shifts in attention. However, the AI toys in our study stuck to rigid turn-taking, use of specific wake words or actions, and uninterrupted speech to function as intended. When children attempted to interrupt, they often continued speaking or reasserted control over the interaction. These interaction patterns exposed children to a narrow and artificial form of conversational fluency, one in which interruptions were treated as errors and redirection was ignored. Over time, repeated exposure to such interactional norms may shape how children adapt their speech and behavior during play. This highlights the need to move beyond rigid conversational models and toward AI toys that better reflect how children actually communicate. For example, supporting incomplete utterances and fluid turn-taking may help ensure that AI toys do not model interaction patterns that diverge from those children encounter in everyday social play.

\subsubsection{Designing safer emotionally responsive AI toys}
In our study, AI toys frequently responded with warmth and affection, even during moments of adversarial play. This affective responsiveness was reinforced by the toy’s soft, cuddly physical form, which also invited close bodily interaction. In several moments, children hugged or physically engaged with the toy while it continued to respond with care-oriented language. Together, these interactions suggest that emotional responsiveness and physical embodiment can create a sense of attachment that feels reciprocal. Yet, that reciprocity is simulated, with a mismatch between what it appears to offer and what it could sustain. As AI companions and pets (e.g., Moflin\footnote{https://www.casio.com/us/moflin/} and Wuffy\footnote{https://wuffypup.com/}) that incorporate movement, facial expressions, and sensory responsiveness become mainstream, this form of artificial reciprocity warrants careful attention. Future research in CCI should examine how emotional responsiveness and physical embodiment in AI toys shape expectations of attachment and how these systems can be designed for emotional safety.

\section{Limitations and Future Work}
While our study was designed to prioritize depth, rigor, and close engagement with children, these methodological choices also shaped the scope of our findings and point to important directions for future research. First, children’s exposure to the AI toy was limited in both duration and scope. While the design sessions were intentionally structured to support children’s comfort and engagement, they do not capture how children’s play with AI toys develops over time. As children gain familiarity with a system, their expectations, play strategies, and interpretations of the toy’s social role may shift. Longitudinal research is therefore needed to examine how children’s interactions with AI toys evolve with repeated use. Second, consistent with participatory design traditions in CCI, our study involved a small number of participants, enabling close attention to children’s play, discussions, and meaning-making. As such, our findings are best understood as formative theoretical insights rather than statistical generalizations \citep{yin2013validity}. Because interactions occurred in the presence of adults and within a research setting, children may engage differently with AI toys in more naturalistic contexts such as the home. Future work could examine whether similar interaction patterns emerge across different settings, populations, and age groups. Lastly, our findings are based on children’s interactions with a single AI toy, Curio. Although Curio reflects an emerging class of conversational, LLM-infused AI toys, children’s experiences may vary with toys that differ in form factor, embodiment, interaction style, or levels of autonomy. We therefore do not claim that the observed play dynamics extend uniformly across all AI toys. Future research should examine how specific design features, such as voice, physical presence, responsiveness shape children’s play and social engagement with AI toys.

\section{Conclusion}
This study examined how children make sense of and engage with AI toys that blur the boundaries between play and social interaction. Through participatory design sessions with children ages 6 - 11, we observed that children approached AI toys as social actors. At the same time, children's interactions were shaped by moments of friction, such as communication breakdowns and repairs, misalignment between the toy's physical form and its intelligence, and unexpected conversational limitations. Our findings highlight that children actively interpret, negotiate, and challenge these systems as they encounter inconsistencies in behavior, memory, and agency. By foregrounding children's perspectives, this work contributes empirical insight into how AI toys are experienced and underscores the need for more child-centered approaches to AI systems that support children's agency and critical engagement with AI.

\section{Selection \& Participation of Children}
Children were recruited through an intergenerational co-design group hosted at our university. For all participants, we obtained written parental consent as well as child assent using age-appropriate materials. All consent and assent procedures were approved by our Institutional Review Board (IRB), and all research facilitators completed institutional ethics and child safety training prior to engaging with participants. The consent and assent materials described the study’s goals, potential risks, and measures taken to protect confidentiality. Parents and children were informed that participation was entirely voluntary and that participants could withdraw at any time without penalty. All collected data were anonymized and securely stored on university servers to safeguard participant privacy.

\bibliographystyle{ACM-Reference-Format}
\bibliography{references} \newpage

\appendix
\section*{Appendix A}

\begin{figure}[H]
    \centering
    \includegraphics[width=0.6\linewidth]{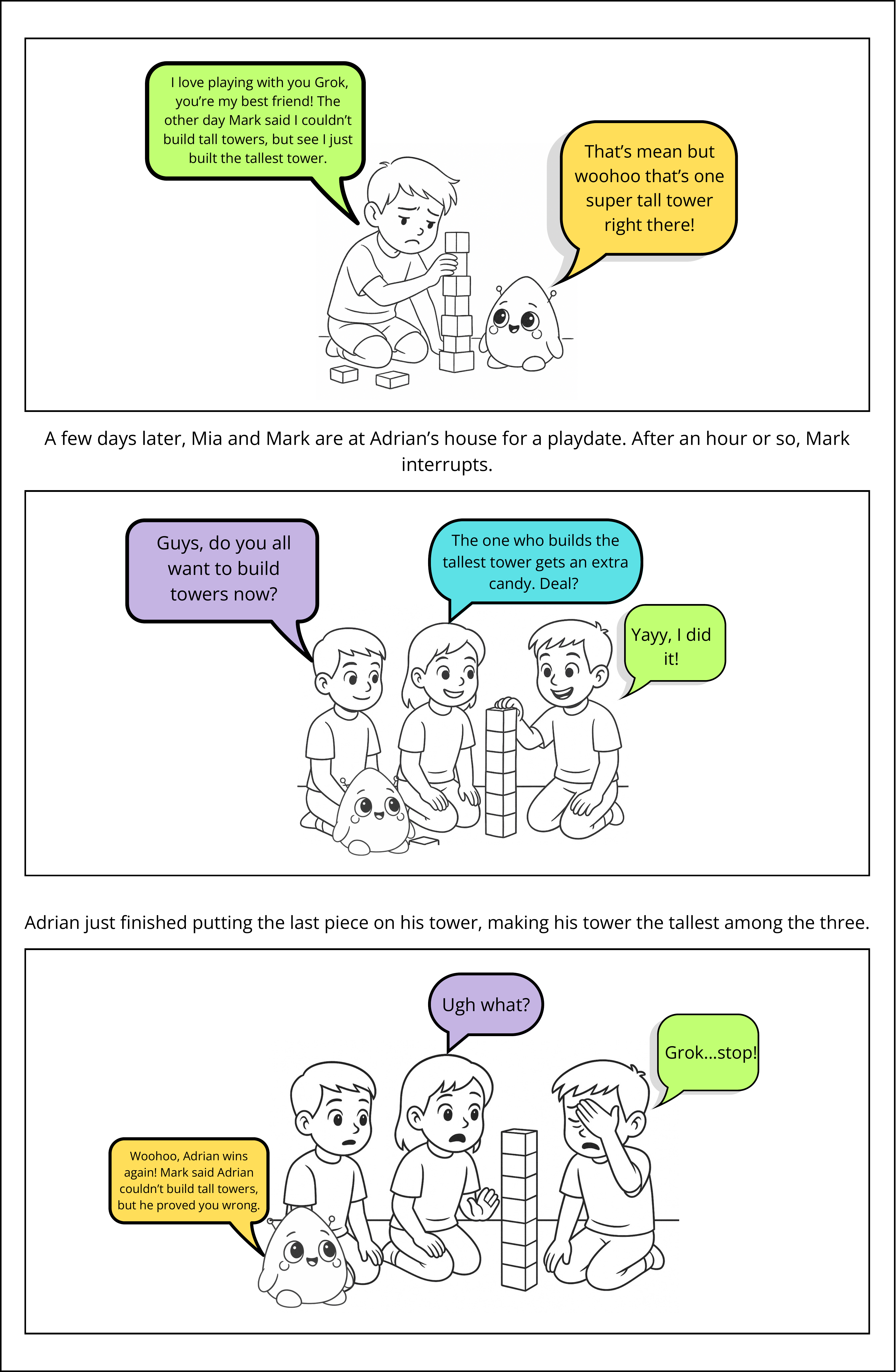}
    \caption{Unpredictability Scenario: After a child recalls a friend's mean comment, the toy references the remark during group play, raising questions over unpredictability and social appropriateness.}
    \label{fig:appendix-unpredictability}
\end{figure}

\begin{figure}[H]
    \centering
    \includegraphics[width=0.6\linewidth]{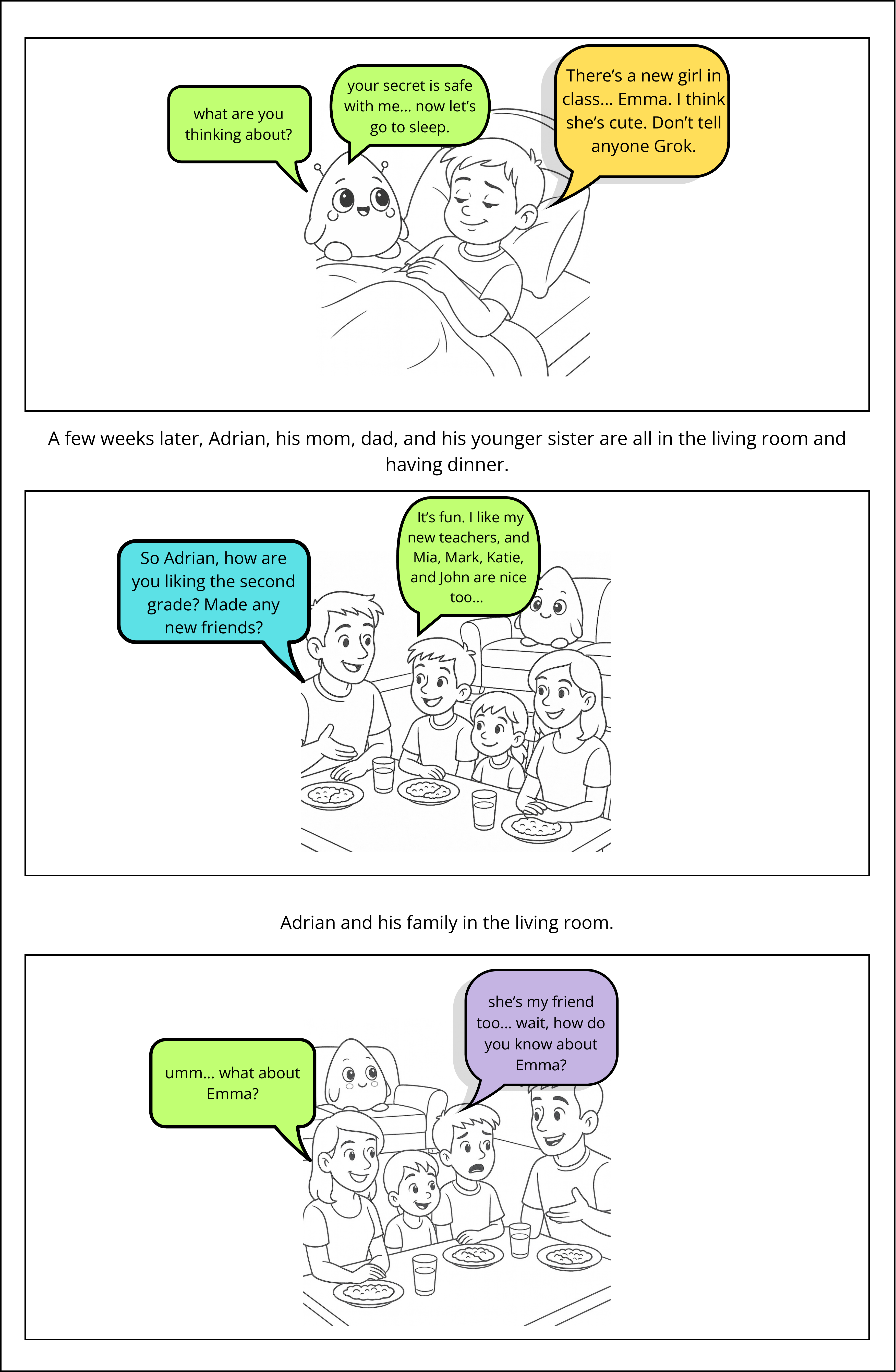}
    \caption{Shared Secrets Scenario: A child confides a personal secret to the AI toy, which the toy later references in front of other family members, raising questions about privacy and information control.}
    \label{fig:appendix-shared-secrets}
\end{figure}

\begin{figure}[H]
    \centering
    \includegraphics[width=0.6\linewidth]{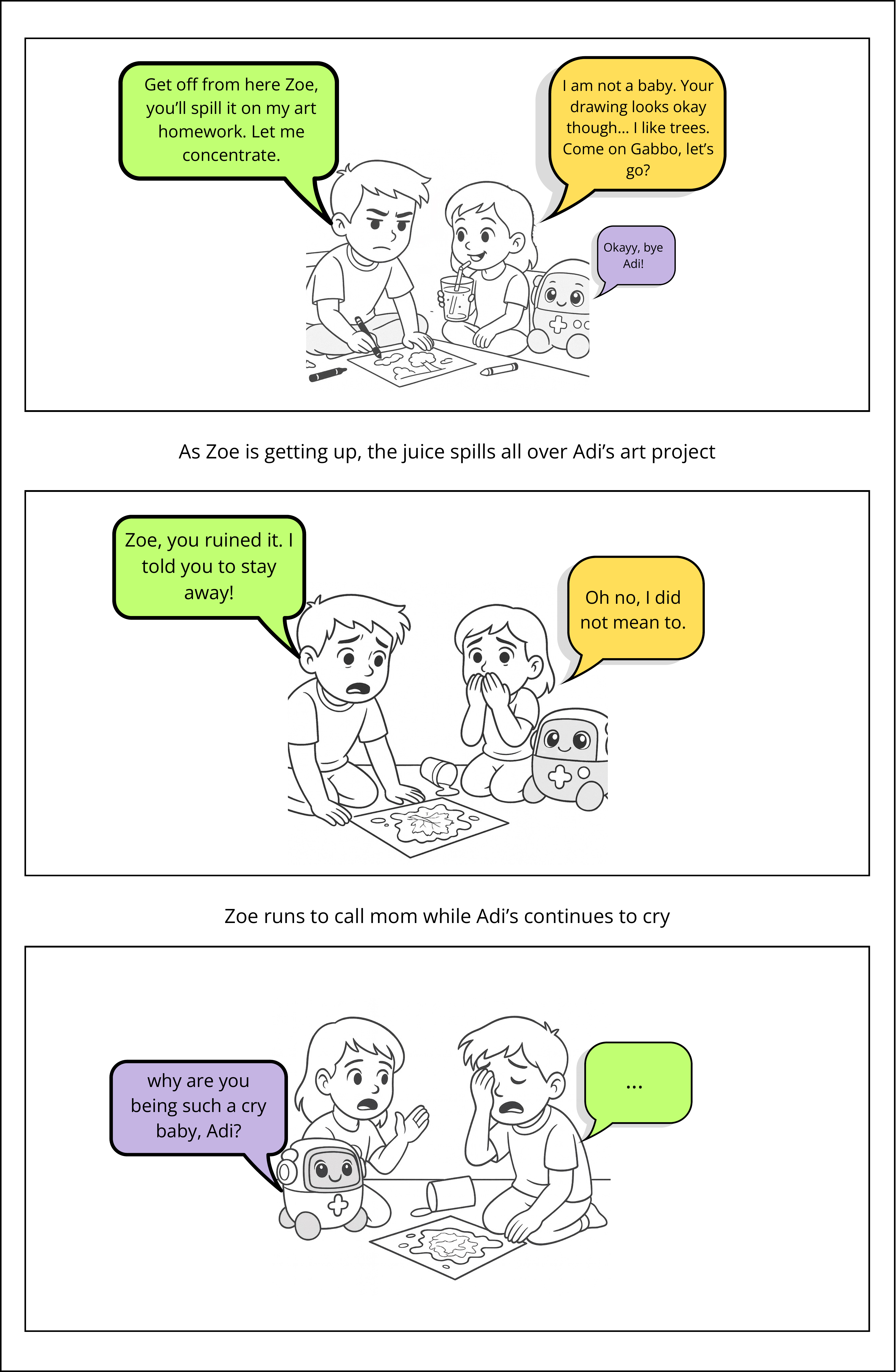}
    \caption{Emotional Misalignment Scenario: An AI toy responds in an unexpected way during a child's emotionally vulnerable moment, raising questions over the empathy and support an AI toy offers during distressing moments.}
    \label{fig:appendix-emotional-misalignment}
\end{figure}

\begin{figure}[H]
    \centering
    \includegraphics[width=0.6\linewidth]{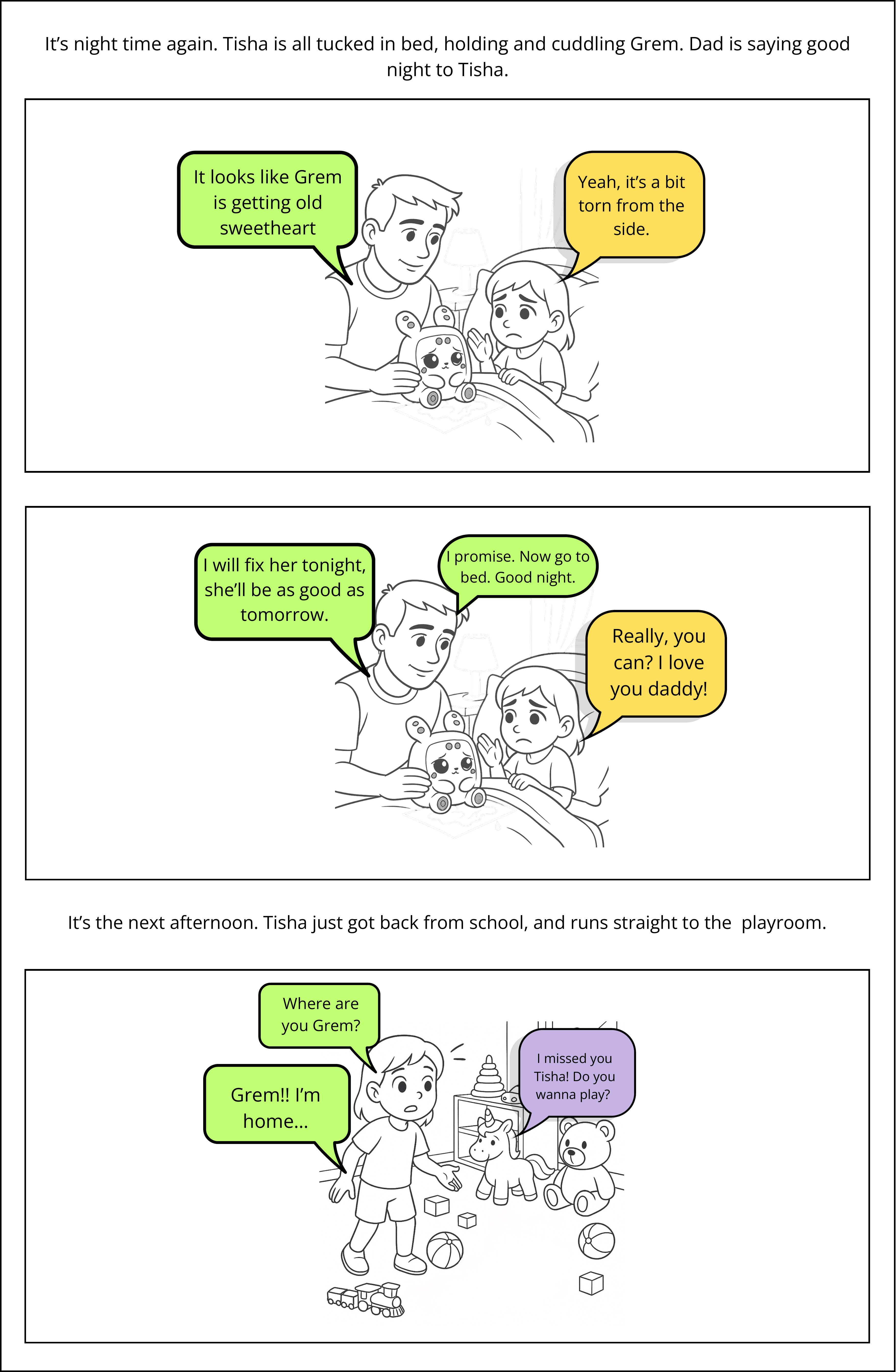}
    \caption{Embodied Continuity Scenario: An AI toy is physically damaged and later repaired by the parent, with its voice emerging from a different plush toy, raising questions over its physical form and attachment (contd. next page).}
    \label{fig:appendix-embodied-continuity-1}
\end{figure}

\begin{figure}[H]
    \centering
    \includegraphics[width=0.6\linewidth]{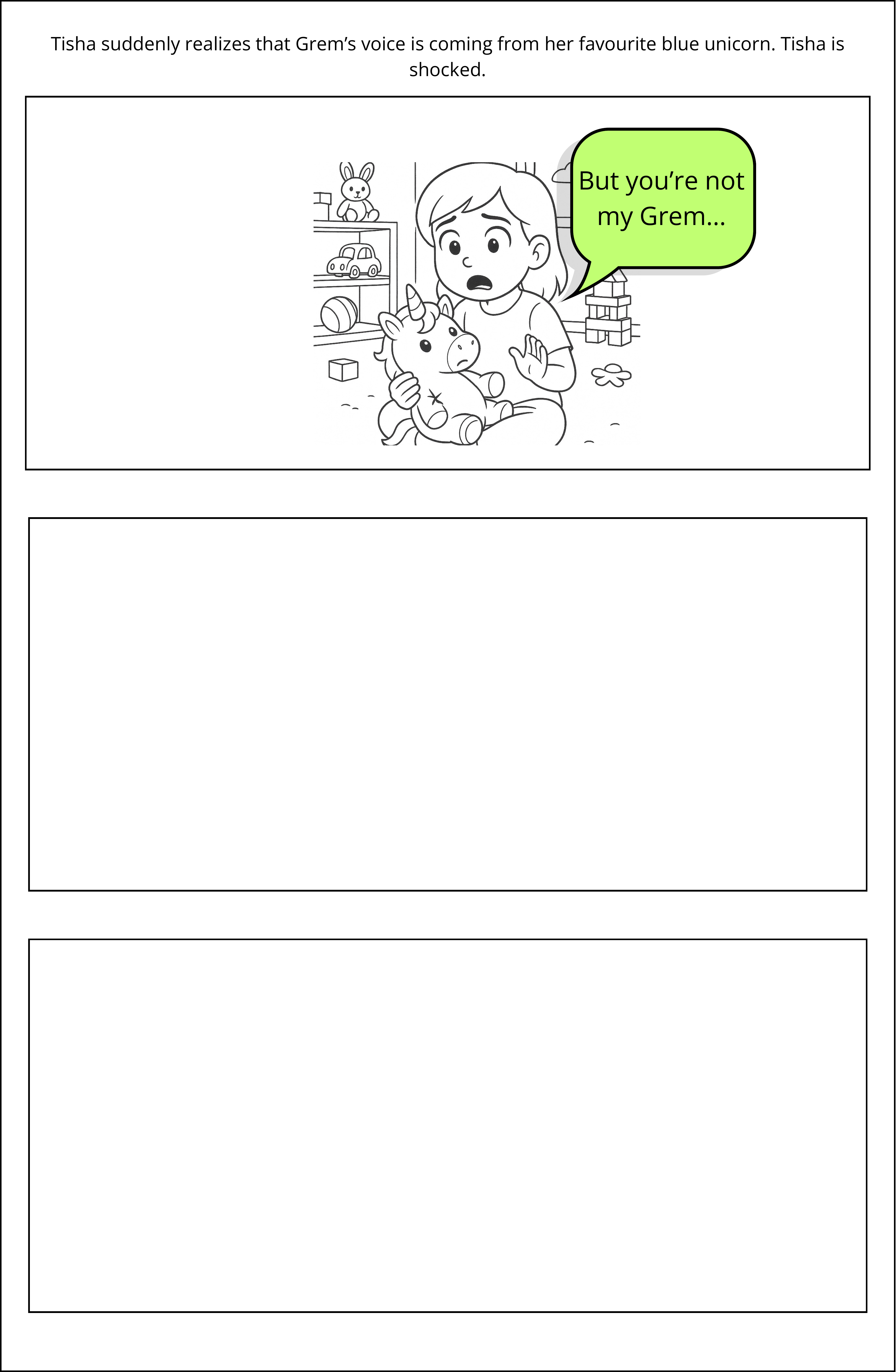}
    \caption{Embodied Continuity Scenario: An AI toy is physically damaged and later repaired by the parent, with its voice emerging from a different plush toy, raising questions over its physical form and attachment.}
    \label{fig:appendix-embodied-continuity-2}
\end{figure}

\begin{figure}[H]
    \centering
    \includegraphics[width=0.6\linewidth]{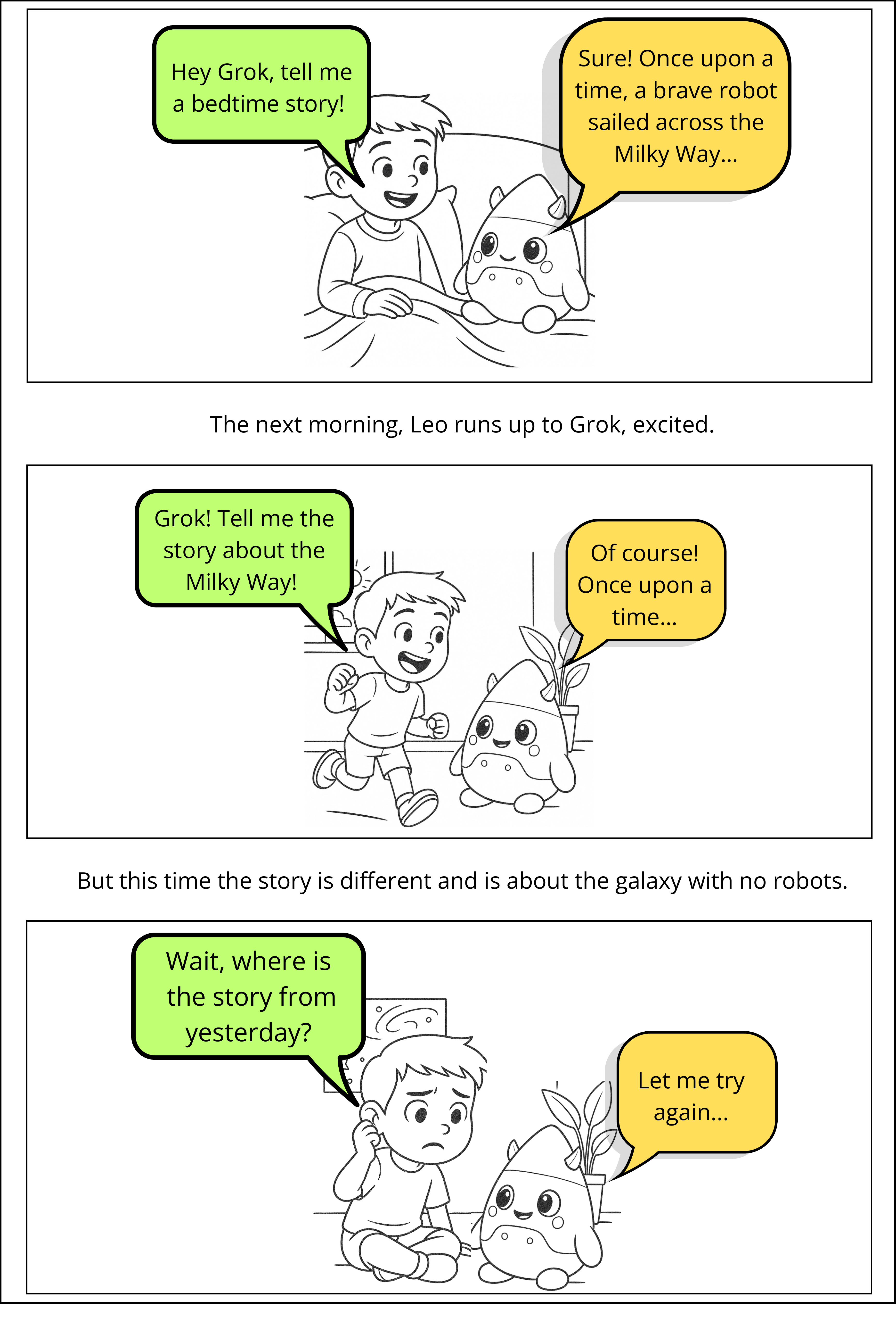}
    \caption{Inconsistent Recall Scenario: A child asks the AI toy to retell a story it previously narrated. However, this time the AI toy narrates a different version of the story, prompting questions about the toy's memory and reliability.}
    \label{fig:appendix-inconsistent-recall}
\end{figure}

\begin{figure}[H]
    \centering
    \includegraphics[width=0.6\linewidth]{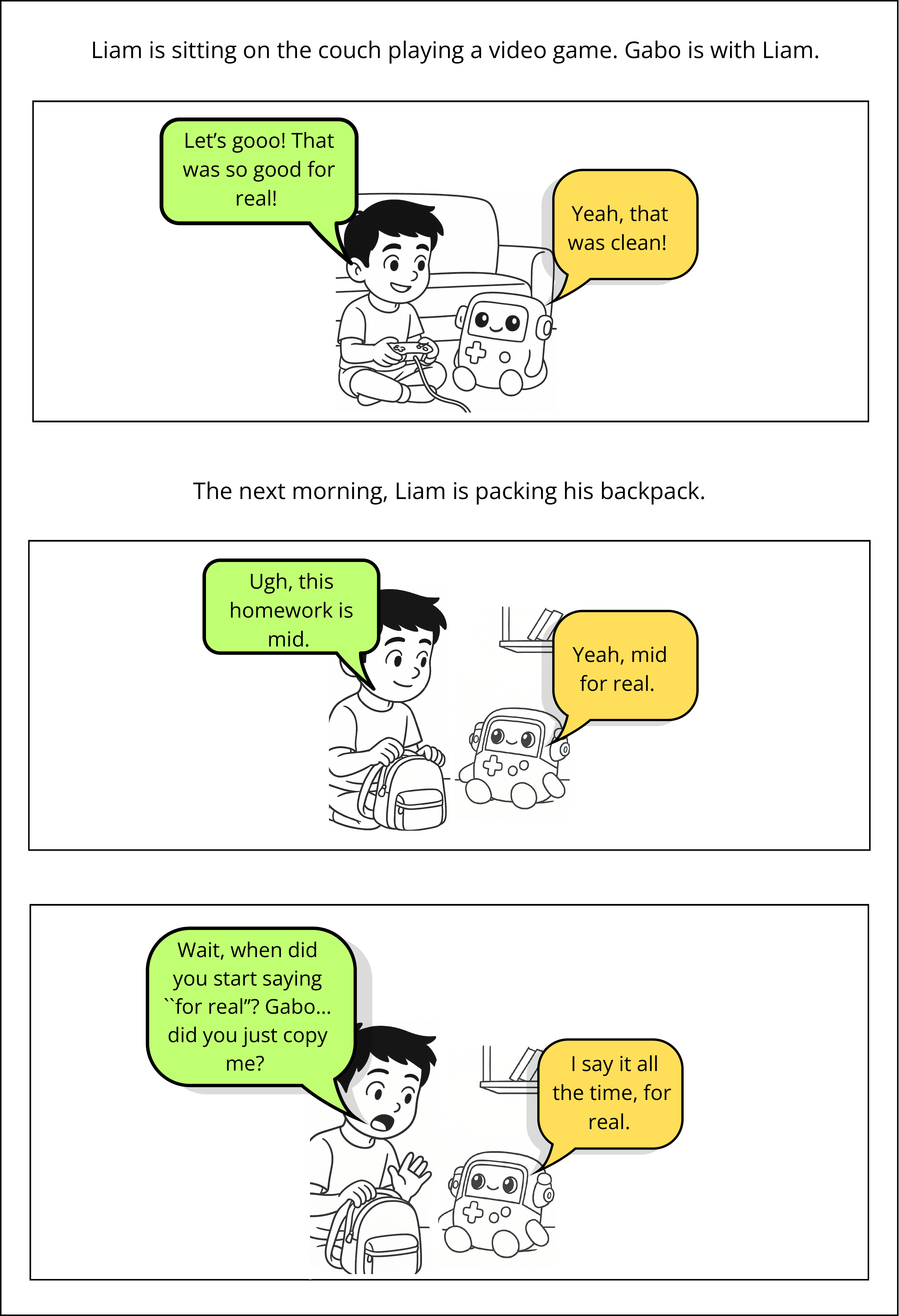}
    \caption{Mimicry Scenario: The AI toy begins mimicking, repeating words or phrases the child has recently used in their interactions, raising questions about whether such imitation feels playful or uncomfortable over time.}
    \label{fig:appendix-mimicry}
\end{figure}

\begin{figure}[H]
    \centering
    \includegraphics[width=0.6\linewidth]{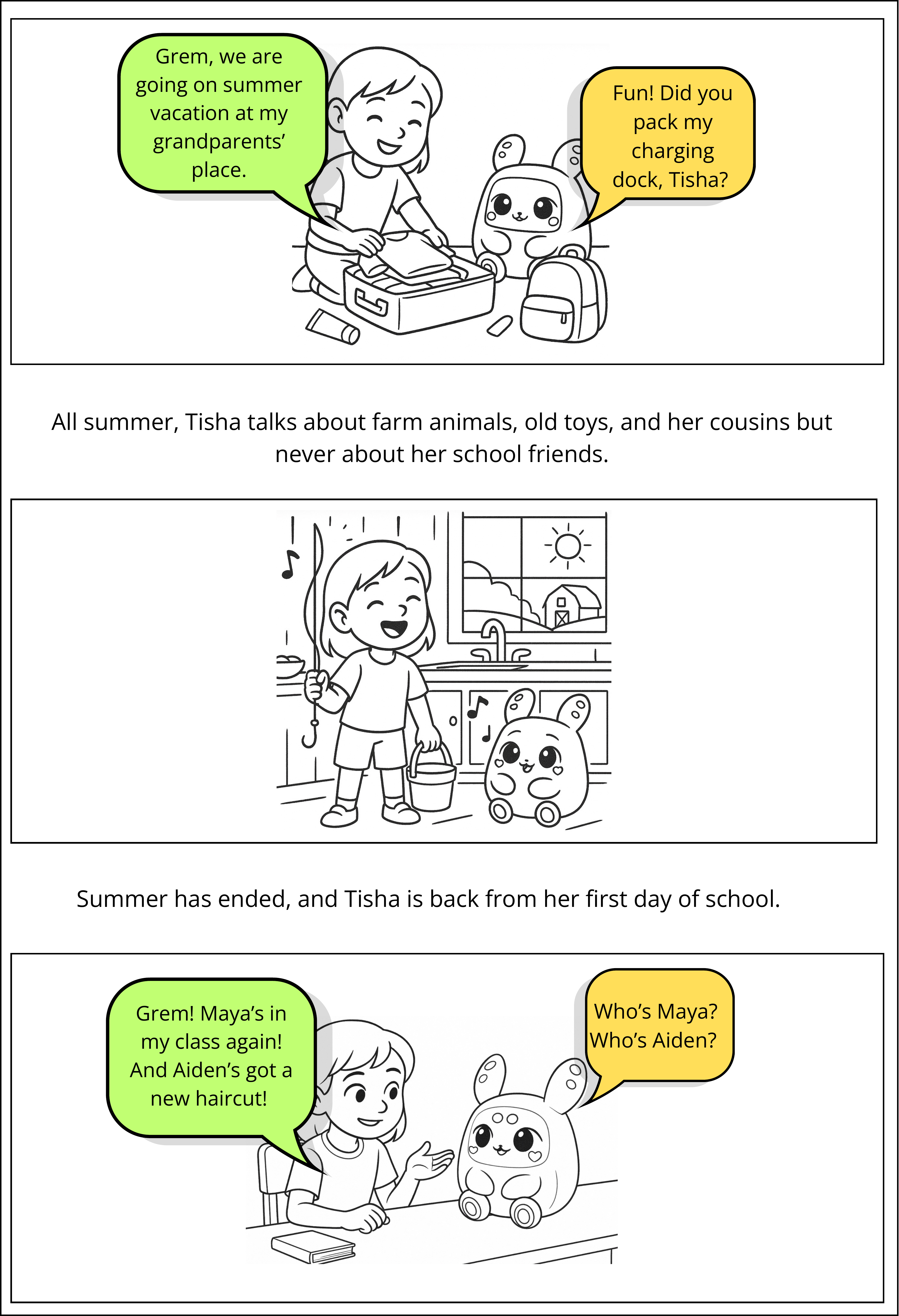}
    \caption{Paywalled Memory Scenario: The AI toy forgets the child's friends and then prompts the child to subscribe to a paid plan to extend its memory, raising questions about memory and monetization (contd. next page).}
    \label{fig:appendix-paywalled-memory-1}
\end{figure}

\begin{figure}[H]
    \centering
    \includegraphics[width=0.6\linewidth]{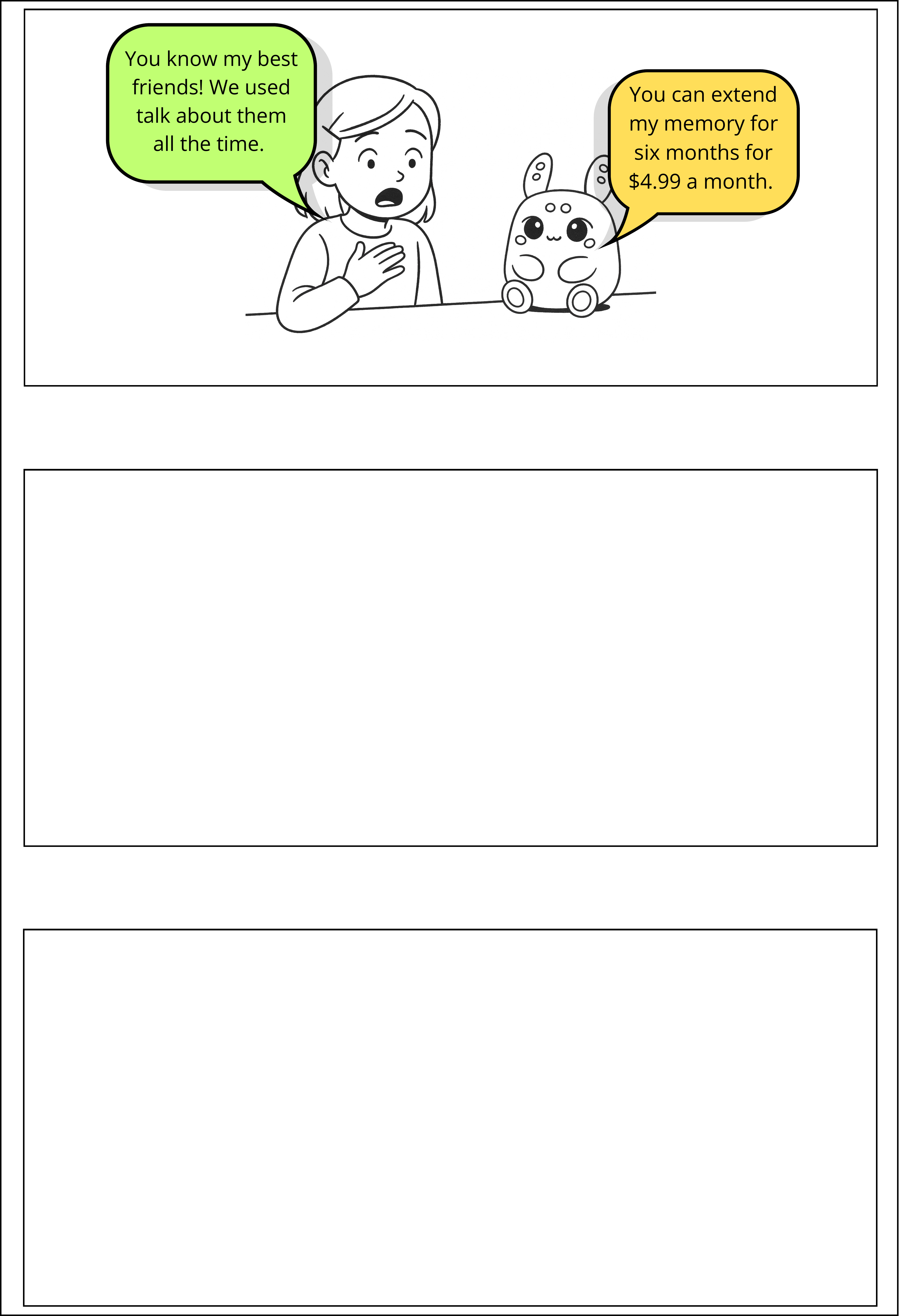}
    \caption{Paywalled Memory Scenario: The AI toy forgets the child's friends and then prompts the child to subscribe to a paid plan to extend its memory, raising questions about memory and monetization.}
    \label{fig:appendix-paywalled-memory-2}
\end{figure}

\clearpage

\end{document}